\documentclass[usenatbib,usegraphicx,useAMS]{mn2e}
\usepackage{amssymb,url,times}
\usepackage{color}

\def\be{\begin{equation}}
\def\ee{\end{equation}}
\def\no{\noindent}
\def\ltsima{$\; \buildrel < \over \sim \;$}
\def\lsim{\lower.5ex\hbox{\ltsima}}
\def\gtsima{$\; \buildrel > \over \sim \;$}
\def\gsim{\lower.5ex\hbox{\gtsima}}

\def\apj{Ap.\ J.}

\def\mnras{Mon.\ Not.\ Roy.\ Astro.\ Soc.}
\def\aap{Astron.\ Astrophys.}

\def\apss{Ap\&SS}
\def\prd{Phys.\ Rev. \ D}

\title[tidal torque in compact binaries]{Tidal torque induced by orbital decay in compact object binaries}

\author[Dall'Osso \& Rossi]{Simone Dall'Osso$^{1}$\footnote{sim.dall@gmail.com} \& Elena M. Rossi$^{1,2}$ \\
$^1$  Racah Institute of Physics, The Hebrew University of Jerusalem, Jerusalem 91904, Israel \\
$^2$ Leiden Observatory, Leiden University, P.O. Box 9513, 2300 RA , Leiden, The Netherlands}
\date{Accepted: 2012, Sept. 19}
\pubyear{2011}

\begin{document}
\label{firstpage}
\bibliographystyle{mn2e}
\maketitle

\begin{abstract} 
As we observe in the moon-earth system, tidal interactions in binary systems can lead to angular momentum
exchange. The presence of viscosity is generally regarded as the
condition for such transfer to happen. In this paper, we show how the orbital evolution can
cause a persistent torque between the binary components, even for inviscid bodies.
This preferentially occurs at the final stage of coalescence of compact binaries, 
when the orbit shrinks successively by gravitational waves and plunging on a timescale shorter than the viscous timescale.
The total orbital energy transferred to the secondary by this torque is $\sim 10^{-2}$ of its binding energy. 
We further show that this persistent torque induces a differentially rotating quadrupolar perturbation.
Specializing to the case of a secondary neutron star, we find that this
  non-equilibrium state has an associated free energy of $10^{47}-10^{48}$ erg, just prior to coalescence.
This energy is likely stored in internal fluid motions, with a sizable amount of differential rotation. 
By tapping this free energy reservoir,
a preexisting weak magnetic field could be amplified up to a
  strength of $\approx 10^{15}$ Gauss.
Such a dynamically driven tidal torque can thus recycle an old
neutron star into a magnetar, with possible 
observational consequences at merger.
\end{abstract}


\section{Introduction}
\label{sec:intro}
The effects of tidal interactions between celestial bodies have been known and studied for 
a very long time (Darwin 1879, Chandrasekhar 1933, Kopal 1959, 1968). These effects range from 
the spin-orbit coupling - as observed in the earth-moon system- to the tidal disruption of astronomical
 objects, as they wonder too close to each other.

In a binary, it is well known that a tidal torque can arise because of viscous 
processes. In this picture, orbital angular momentum is transferred to the stellar spin.
If the process is efficient, the spin frequency may eventually equal the orbital frequency
and the torque will vanish.
This configuration is a called ``tidally locked'' or ``synchronized'' binary. 
This is typically observed in binaries containing normal stars (Zahn 1977).

In two classical papers, Kochanek (1992) and Bildsten \& Cutler (1992) showed that, 
in {\em compact} object binaries containing a neutron star (NS), 
there is not enough energy in the tides to lock a NS spin to its orbital motion, regardless
of the magnitude of viscosity. Indeed, these authors further showed that all possible estimates for the NS viscosity
would give only negligible spin energy.

In this paper, we show that, when the inspiral is driven by gravitational waves, a torque of {\em dynamical} origin can  
ensue, even in absence of viscosity. The magnitude of this torque exceeds that due to viscosity in compact object binaries, at their final stage
of coalescence.  Because of the negligible effect of viscosity, the differential nature of the tidal torque induces a {\it radially dependent} rotating quadrupole perturbation.  
We quantify the total energy and angular momentum carried by this perturbation and its radial structure. We find that it does not correspond to a minimum energy level. 
The free energy available is typically
$10^{47}-10^{48}$ erg, which may have interesting observable consequences.

The paper is organized as follows.
In Sections 2, we show that the dominant tidal
 torque prior to coalescence of two compact objects is that produced by their orbital evolution.
For regular stars, instead, the tidal torque is produced by viscous processes only.
In \S 3 we derive the tidal radius and the radius where the dynamical instability sets in. In \S 4 we summarize the 
ordering of the characteristic radii, for different binary systems.
In \S 5, we calculate the total amount of energy in the tides. In \S 6, we show the radial dependence of the 
rotating quadrupole excited in the secondary. In \S 7, we specialize to the case of a neutron star, described by a
polytropic equation of state and we quantify the free energy stored in the rotating quadrupole. Finally, in \S 8, we 
discuss a possible outcome of this free energy and we draw our conclusions.
\section{The integrated tidal torque onto the secondary}
\label{sec:Nt_integral}

As it was shown in classical works dating back to Darwin (1879) and Chandrasekahr (1933), tidal interactions
in binaries can exchange angular momentum between the orbit of the system and each individual star.
Let's denote $M_1$ the mass of the primary star, $M_*$ the mass of the secondary and $R_*$ its radius. 
Tidal forces cause a departure from sphericity, which in the secondary's local frame 
corresponds to a non-resonant rotating quadrupolar tide (Thorne 1998). For brevity, we will frequently refer to this
quandrupolar perturbation simply as  the {\it tidal bulge}. 

In this paper, we work in the approximation of small tidal deformations,
which are valid for orbital separations $a$ larger than the tidal radius 
$$ a_{\rm T} \propto ~R_* /q^{1/3},$$
where $q$ is the mass ratio $q = M_*/M_1$. 
The numerical coefficient is $\approx 2$ and it will be calculated as a function
of the stellar internal structure in in Sec. \ref{sec:tidalradius}.

If the tidal deformations are not perfectly aligned with the line joining the star centres (thereafter ``line of centres" in short), 
each component will exert a non-zero torque on the companion and a flux of angular momentum is set up in the system. 
In the presence of a finite viscosity, there is a coupling between the orbital angular momentum and the stellar {\it spin}.
For compact object binaries, however, this viscous coupling is {\it not} efficient, and the angular momentum exchange 
is between the orbit and the {\it stellar bulge} only (see section~\ref{subsec:fix_separation}).

\begin{figure}
\includegraphics[width=0.95\columnwidth]{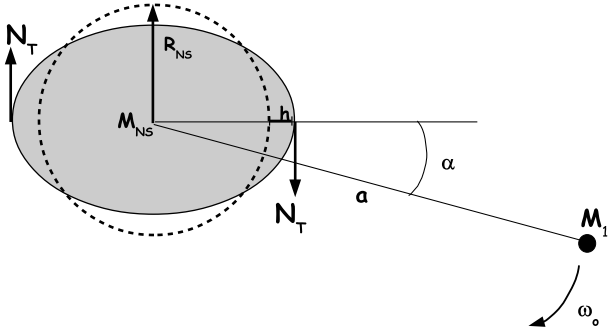}
\caption{Sketch of the tidal torque which arises in a binary, when a tidal bulge is misaligned. 
The torque is visualized in the orbital plane, in the frame of the secondary (the larger, gray star). 
In this frame, the primary (the small black star)
rotates about the secondary at the orbital frequency (we consider non-spinning stars).
At the location of the secondary the dashed line indicates the contour of an unperturbed, spherical star.}
\label{fig:tide_1}
\end{figure}

We sketch here the classical derivation of the total tidal torque ($N_{\rm T}$) applied onto the
secondary, when its bulge is misaligned by an angle $\alpha$ (see Chandrasekhar 1933, Alexander 1973 and references therein). 
We assume small lag angles throughout the paper and, when needed, we will expand the trigonometric functions accordingly, without explicitly mentioning it.
We will show that the small angle approximation is justified  for $a \ge a_{\rm T}$ and for typical stellar dynamical viscosity.

We consider a two dimensional problem, in the orbital plane of the binary (see Fig.1).
There, the primary at a distance $a$ generates a tidal field $\psi_{\rm T}$ at
the location of the secondary given by (Alexander 1973 and references therein)

\be
\psi_{\rm T} = \frac{G M_1}{a} \left(\frac{r}{a}\right)^{2} \frac{3\cos^2(\theta)-1}{2},
\label{eq:psit}
\ee

\no
where the polar coordinates (r, $\theta$) are centered in the centre of mass of the secondary, and
$\theta$ is measured from the semi-major axis of the bulge. In eq. \ref{eq:psit}, $0 \le r \le R_*$.

The secondary reacts to the tidal pull readjusting to a new hydrostatic configuration on a timescale
$t_* = 2 \pi/\omega_*$. This mass redistribution generates
in turn a non-spherical perturbation $\psi_{*}$ to the secondary's gravitational potential at each external point $r > R_*$,

\be
\psi_{*} = \kappa_2 \left(\frac{R_*}{r}\right)^3 \times \psi_{\rm T},
\label{eq:psistar}
\ee

\no
where $\kappa_2$ --the Love number-- measures the global deformability of the
secondary (Love 1909, Hinderer 2008).
%{\bf give reference}.
Because of $\psi_*$, the secondary will apply a torque onto the primary given by 
$\bf{\nabla_{\theta}\psi_{*}}(r=a) \times \bf{a}$. This is equal, but opposite in direction, 
to the tidal torque that the primary exerts onto the secondary. 
Therefore,
\begin{equation}
\label{classicaltorque}
N_{\rm T} = \frac{3 \kappa_2}{2} \frac{G M^2_{\rm 1}}{a}
\left(\frac{R_*}{a}\right)^5 \mbox{sin}2\alpha \simeq \frac{3 k_2 \omega_*^2 M_*
  R^2_* \alpha}{(1+q)^2}\left(\frac{\omega_{\rm o}}{\omega_*}\right)^4.
\label{eq:Nt}
\end{equation}
The  value of $\alpha$ in the above expression must be regarded
as an {\em effective} lag angle: a global property  of a star with
radius $R_*$ and total deformability $\kappa_2$.

The torque vanishes when either the secondary or the secondary's tidal bulge rotates with an angular frequency equal to the
orbital frequency (i.e. $\alpha=0$). In the following sections, we discuss the conditions for this synchronization {\it not} to happen
and its physical consequences.

\subsection{The tidal interaction of a binary at a {\em fixed} separation}
\label{subsec:fix_separation}

\subsubsection{the inviscid case}
Consider the simplest case first: a binary at a fixed separation $a$, and a non-spinning secondary. 
We work in a non rotating frame centered onto the secondary.  

In the absence of orbital motion, the tidal bulge is formed in a timescale $t_*$ and it lies perfectly 
aligned with the primary's position.
As soon as orbital motion is switched on, the primary begins to rotate about the secondary at the
orbital frequency $\omega_{\rm o} = \sqrt{GM_{\rm tot}/a^3}$, where $M_{\rm tot} = M_1 + M_*$.
The bulge is initially at rest for a reaction time $t_*$. In this time interval, the bulge lags behind the 
line of centres by a linearly increasing angle $\alpha \simeq \omega_{\rm o} t$.
As $\alpha$ grows, 
a tidal torque ($N_{\rm T}$) ensues that transfers angular momentum to the bulge. This process will last
until the bulge acquires the angular momentum $J_{\rm b} \equiv I_{\rm b} \omega_{\rm b} = I_{\rm b} \omega_{\rm o}$, and it co-rotates with the primary. Here $\omega_{\rm b}$ 
and $I_{\rm b}$ are the bulge rotation frequency and \textit{``effective"} moment of inertia, respectively. 
Thus we can write $N_{\rm T} t_* \simeq J_{\rm b}$.

Once this angular momentum has been transfered, the bulge rotates at $\omega_{\rm b} = \omega_{\rm o}$ and
it is perfectly aligned with the line of centres. The torque vanishes again, having reached a stationary 
configuration\footnote{The angular momentum $J_{\rm b}$ is taken from the orbit through the action of the 
torque $N_{\rm T}$. This causes the orbit to loose a tiny amount of angular momentum, thus slightly shrinking and spinning up. 
However the orbit has an effective moment of inertia $I_{\rm o} \gg I_{\rm b}$ (as we show below). 
It is thus clear that $N_{\rm T}/I_{\rm o} = \dot{\omega}_{\rm o} \ll N_{\rm T}/I_{\rm b} = \dot{\omega}_{\rm b}$. Therefore, the bulge frequency quickly 
reaches the orbital frequency and the torque drops accordingly to zero. 
This mechanism is self-regulating and does not produce any instability of the system.}. 
From the above we can derive an expression for the effective moment of inertia of the bulge 

\be
I_{\rm b} = t_* N_{\rm T}/\omega_{\rm o} = 6 \pi^2 \kappa_2/(1+q)^2 (\omega_o/\omega_*)^4 M_* R^2_*,
\label{eq:Ib}
\ee
where we used the fact that  $\alpha$ grows linearly in time during this transient phase.
Eq.~\ref{eq:Ib}  shows that $I_{\rm b}$ is much smaller than the stellar moment of inertia $I_* \approx M_* R^2_*$, given its 
strong dependence on the ratio $(\omega_o/\omega_*)$ which is by definition $\ll 1$.
The angular momentum that needs to be transferred to the tidal bulge for it to co-rotate is then 
$J_{\rm b} = I_{\rm b} \omega_{\rm o} \propto M_* R^2_* \omega_o (\omega_o / \omega_*)^4$.

\subsubsection{the effect of  viscosity}
\label{subsec:viscosity_fixr}
In the ideal case, where the secondary is perfectly inviscid, the circulation energy and the angular momentum of the bulge 
stay constant once the co-rotation is reached, and there will always be alignement of the bulge with the line of centres.
In fact, the secondary has a finite - though {\it small} - viscosity,
which implies a continuous transfer of angular momentum from the tidal bulge to the stellar spin. 
Thus, the tidal bulge is spun up by the primary torque but simultaneously slowed down by the viscous torque.
The net result is that the bulge cannot keep pace with the orbital motion of the primary and it lags behind the line of centers 
by an angle $\alpha_{\rm v}$. This, in turn, implies the existence of a persistent tidal torque. 
Angular momentum is thus continuously transferred from the orbit to the bulge by the tidal torque and from the bulge to the secondary's 
spin by viscous stresses. At equilibrium
the two transfer rates are equal: $N_{\rm T} =J_{\rm b}/t_{\rm v}$, where $t_{\rm v} = R_*^2/\nu$ is the characteristic viscous timescale determined by the kinematic viscosity
$\nu$. This condition determines the equilibrium value of the lag angle,
\begin{equation}
 \alpha_{\rm v} \approx  2 \pi^2 \omega_o /(\omega^2_* t_{\rm v}).
\label{eq:av}  
\end{equation}
where we used eq.~\ref{classicaltorque} and eq.~\ref{eq:Ib}, and the fact that $\alpha_{\rm v}$ is small to write $\omega_{\rm b} \sim \omega_{\rm o}$.

The above derivation of $\alpha_{\rm v}$ for weak viscous stresses is our alternative derivation 
to the classical one, which considers the response to the tidal field of each fluid element in the 
secondary star (cfr. Kopal 1968, Alexander 1973, Cutler \& Bildsten 1992 and references therein). 
The fluid element will feel a time-varying external potential, the tidal potential of the primary, 
oscillating at twice the orbital frequency, $2 \omega_{\rm o}$. This drives small radial displacements $x(t)$ 
of the element from its equilibrium position. The fluid element is also subject to a restoring force, 
the self-gravity, which has a typical oscillation frequency $\omega_*$. Finally, a weak friction, 
$t_{\rm v} \gg  t_*$, is contrasting its oscillation. The equation of motion of each fluid element 
is thus that of a forced, (under)damped oscillator

\begin{equation}
\label{eq.ho_v}
\ddot{x}(t) + 2 t_{\rm v}^{-1} \dot{x}(t) + \omega^2_* x(t) = A \,\mbox{e}^{i 2 \omega_{\rm o} t}.
\end{equation}

The solution is a displacement $x(t) \propto e^{i(2 \omega_{\rm o} t - \alpha_{\rm v})}$ 
which lags in phase with respect to the driving forcing $\propto$ $e^{i 2 \omega_o t}$
by an angle $\alpha_{\rm v} \propto \omega_o/[(\omega^2_*-\omega_{\rm o}^2) t_{\rm v}] \sim \omega_o/(\omega^2_* t_{\rm v})$.
Thus, we recover our result (eq.~\ref{eq:av}).

If we start with the system at rest, the equilibrium value for the lag angle, $\alpha_{\rm v} \approx \omega_o t_{\rm lag}$,
is reached after a transient linear growth on a timescale $t_{\rm lag} \approx t_* (t_*/t_{\rm v}) \ll t_*$. This confirms our 
picture that, in the presence of viscosity, the system reaches first an equilibrium in which it does {\em not} perfectly co-rotate.
In fact, it lags behind by 
a small but finite angle 
\be
\alpha_{\rm v} = (\omega_{\rm o} - \omega_{\rm s}) t_{\rm lag},
\label{eq:alpha_v_tlag}
\ee
where $\omega_{\rm s}$ is the secondary's spin frequency. 

\subsection{The effect of orbital shrinking}
\label{sec: orbitalshrinking}
In the previous section, we considered a fixed orbital separation. However, binaries may lose orbital energy and angular momentum
 by emission of gravitational waves. As a consequence, they will be brought towards the plunging phase which leads to coalesce through a succession of circular orbits on a timescale

\be
\label{taugw}
t_{\rm GW} \approx  1.2~{\rm yr}~\left[\frac{\omega_{\rm o}}{1~{\rm rad/s}}\right]^{-8/3} \frac{(1+q)^{1/3}q^{2/3}} {m^{5/3}_*},
\ee
where $m_* = M_*/M_{\sun}$ (Peters 1964).

\no
Indeed, for solar mass binaries which have a separation smaller than $a \approx 3 R_{\sun}  $, $t_{GW}$ is less 
than the age of the universe and merger can occur. However, for regular stars, this separation is smaller than 
their tidal radius, and tidal disruption inevitably occurs before GWs can become efficient in removing orbital 
angular momentum. Therefore, to investigate the effect of orbital decay by GW emission, we should consider hard 
binaries, hosting {\it compact} objects. Their orbital shrinking changes substantially the classical picture of 
tidal interactions in binaries depicted earlier. 

\subsubsection{the inviscid case}
We first ignore viscosity in the secondary star and consider, as a starting
configuration, a 
%perfectly synchronized 
binary in which the secondary's tidal bulge rotates exactly at the orbital frequency.
As the orbit shrinks and, thus, the orbital frequency increases, the tidal 
bulge finds itself lagging behind and therefore subject to a tidal torque. This torque tries to  spin it up to track the evolving orbital motion of the 
primary. As discussed above, the spin-up of the bulge occurs on a timescale $\sim t_*$. 

The lag angle induced by the orbital shrinking, $\alpha_{\rm GW}$, can be quantified in different ways.
The easiest is to calculate the angle travelled by the primary on a time
$t_*$, as its orbital frequency grows at a rate $\dot{\omega}_{\rm o}$, while the bulge lags behind at the original 
$\omega_{\rm o}$: $\alpha_{\rm GW} = (1/2) \dot{\omega}_{\rm o} t^2_* \approx \omega_{\rm o}/(\omega^2_* t_{\rm GW})$.

For an alternative, insightful derivation of $\alpha_{\rm GW}$, we consider the dynamics of the system.
The orbit loses angular momentum to GW emission, and we call the associated torque $N_{\rm GW}$. 
As a consequence, the orbit spins up and a tidal torque $N_{\rm T}$ arises.
 This torque subtracts further orbital angular 
momentum, transferring it to the secondary's bulge. Equilibrium is reached when the bulge frequency 
becomes able to change at the same rate as the orbital frequency, $\dot{\omega}_{\rm o} = \dot{\omega}_{\rm b}$.
This requires the existence of a non-zero tidal torque -- and thus of a non-zero 
lag angle $\alpha_{\rm GW}$--, which continuously spins up the bulge at the appropriate rate, 

\begin{equation}
\label{equilibrium_GW}
\frac{N_{\rm T}}{I_{\rm b}} = \frac{N_{\rm GW}+N_{\rm T}}{I_{\rm o}},
\end{equation}

\no
where $I_{\rm o}$ is the binary moment of inertia. Since the orbital decay is largely 
driven by GW emission, we neglect the second term on the right-hand side of equation \ref{equilibrium_GW} 
and solve for the equilibrium angle 

\begin{equation}
\label{eq:agw}
\alpha_{\rm GW} = 2 \pi^2 \frac{\omega_{\rm o}}{\omega^2_* t_{\rm GW}}.
\end{equation}

Finally, in analogy with section \ref{subsec:viscosity_fixr}, we can derive $\alpha_{\rm GW}$ from the oscillatory motions 
of fluid elements in the secondary star, which are forced at an increasing orbital frequency. Differently from eq.~\ref{eq.ho_v}, 
we neglect here the damping term but allow for a time-dependent forcing term, 

\begin{equation}
\label{alpha_gw_oscillator-N1}
\ddot{x}(t)+\omega^2_* x(t) =  C e^{ i \int{2 \omega_{\rm o}(t) dt}}.
\end{equation}
\no
The solution for the radial displacement is  $x(t) \propto e^{i\left[\int{2 \omega_{\rm o}(t) dt}-\alpha \right]}$.
Therefore, even in the absence of viscosity the evolving orbital frequency will cause the displacement of each fluid element to lag in phase
with respect to the perturbing potential by the angle $\alpha$. To first order in $(\omega_{\rm o}(t)/\omega_*)^2$, we get
$\alpha \equiv \alpha_{\rm GW} \propto \omega_{\rm o}/ [(\omega^2_* - \omega^2_{\rm o}) t_{\rm GW}] \sim \omega_{\rm o}/ (\omega^2_*t_{\rm GW})$.

From these arguments, we conclude that, even in a perfectly inviscid secondary, the tidal bulge will develop an angular lag 
and will thus be continuously subject to the minimal tidal torque associated to it (Lai, Rasio, Shapiro 1994b, Lai \& Shapiro 1995).

\subsubsection{In the presence of viscosity}
In the presence of viscosity, there is a net transfer of angular momentum from the tidal wave to the secondary's spin. 
In section \ref{subsec:viscosity_fixr} we showed that, for a given $\omega_{\rm o}$, the value of $\alpha_{\rm v}$ is determined by the balance between 
the viscous torque, draining angular momentum from the rotating bulge, and the tidal torque, that acts to replenish it.
If there {\it is} orbital shrinking, but it proceeds at a slow pace, i.e. $t_{\rm GW} \gg t_{\rm v}$, the effect of $\dot{\omega}_{\rm o}$ can be neglected
and the picture of section \ref{subsec:viscosity_fixr} is still the valid description of the equilibrium configuration.
If, on the other hand,  $t_{\rm GW} \ll t_{\rm v}$,
the lag angle is instead determined by the orbital evolution as described in the previous section.
Note that $t_{\rm GW} \ll t_{\rm v}$ implies $\alpha_{\rm v} \ll \alpha_{\rm GW}$ and viceversa (compare eq.~\ref{eq:av} and eq.\ref{eq:agw}). This means
that during the binary evolution the effective tidal torque is the maximum between the ``viscosity-induced" and the ``GW-induced" torque.

Since $t_{\rm GW} \propto a^4$, we expect the GW-induced torque to be important at small binary separation.
Indeed, $\alpha_{\rm GW}  > \alpha_{\rm v}$ when

\begin{equation}
\label{eq:wo_lim}
\omega_o > 1.1~ m_1^{-5/8} \left(\frac{t_v}{\mbox{yr}}\right)^{-3/8} \frac{(1+q)^{1/8}}{q^{3/8}}~\rm{rad \, s^{-1}},
\end{equation} 
 where $m_{\rm 1} = M_{1}/M_{\sun}$.
In ordinary stars, the effective viscosity operates on a timescale of $\approx$ yr (Zahn 1977, Hut 1981, Goldreich \& Nicholson 1989). 
Therefore, an equal mass binary with a solar type companion
will never enter the high frequency regime of eq.~\ref{eq:wo_lim}, since it requires $a< 6 \times 10^8$ cm $\ll R_{\sun}$. 
The estimated viscous timescale in a typical NS ranges from $t_{\rm v} \sim 0.1$ s for the crust anomalous viscosity (Kochanek 1992),
to $t_{\rm v} \approx 1$ yr for shear viscosity (cfr. Andersson 2007). For WDs 
the shear viscosity acts on a very long timescale 
$t_{\rm v} \sim 10^{12} \div 10^{14}$ s (cfr. Neubauer 1986). 
The internal magnetic field has been proposed as a possible source of anomalously high shear viscosity 
in WDs, giving a somewhat shorter $t_{\rm v} \approx 10^6$ s.
These ranges of timescales together with the large compactness of these objects allow for the condition eq.~\ref{eq:wo_lim} to be satisfied
for  binary separations {\em much} larger than their tidal radius. 
This can be easily seen by recasting eq.~\ref{eq:wo_lim} in terms of $a$,

\begin{equation}
\label{eq:a_lim}
\frac{a}{a_{\rm T}} < 10^4 m_1^{-2/3} m_*^{1/3} \left(\frac{12\mbox{~Km}}{R_*}\right) \left(\frac{t_v}{\mbox{yrs}}\right)^{1/4}.
\end{equation}

\section{Tidally-induced dynamical instability}
\label{sec:instability}
When the orbital separation in a binary becomes comparable to the size of its 
components, tidal effects cause a non-negligible steepening of the interaction 
potential with respect to the term $\propto 1/a$. This increases the radial 
component of the gravitational force, and alters the
keplerian relation between orbital separation and period. Eventually, a point
can be reached where no closed orbits are possible and the system becomes 
dynamically unstable. The binary is brought to coalescence in a 
dynamical (free-fall) time, much faster than dictated by GW emission (Lai \&
Shapiro 1995 and references therein). The same arguments that lead to the  
definition of $\alpha_{\rm GW}$ thus lead to the definition of a new lag angle, 
$\alpha_{\rm dyn}$. This is obtained from the former, by simply 
substituting $t_{\rm GW}$ with the dynamical time in the relevant 
regime.

 In order to determine when the above instability occurs,
it is first necessary to calculate the star deformation as a function of 
the orbital separation. In practise, we calculate the evolution 
of $\eta \equiv h/R_*$, the fractional hight
of the tidal bulge, which allows us to derive the 
perturbed stellar radius,
$$R(a) = R_* [1+\eta(a)]. $$
In fact, this is the semi-major axis of the star: the elongation along the line of centres ($\theta=0$).
To derive $\eta$, we will impose (in Sec. 3.1) that at each radius the secondary's total energy is 
equal to that for a point mass (the tidal term causes just a redistribution of energy).
This calculation will also 
allow us to find an analytical formula for the tidal radius $a_{\rm T}$ and determine 
the radius, $a_{\rm c}$, at which the binary comes into physical contact.
Then (Sections 3.2 and 3.3),  we will write the total (newtonian) effective potential for the binary system
and find at which separation a minimum cannot be found anymore.
This separation $a_{\rm d}$ marks the onset of the dynamical instability.
In all cases, our conclusions compare well with published results
 (cfr. Clark \& Eardley (1977), Lai \& Shapiro (1995), Bini, 
Damour \& Faye (2012) and references therein).

\subsection{Tidal stellar deformation and tidal radius}
\label{sec:tidalradius}

The secondary's tidal bulge is a result of the work done by 
the first term in the expansion of the primary's 
newtonian potential $\psi_{\rm T}$ (eq.\ref{eq:psit}). This causes a decrease in the binding energy of the originally unperturbed 
secondary. In turn, the change in the secondary's external potential, $\psi_*$ (eq. \ref{eq:psistar}), is a 
result of the tidal bulge: it increases the binding energy of the orbit, 
further drawing from the stellar binding energy.
 In the system's energy budget,
the additional energy terms $\psi_{\rm T}$ and $\psi_*$ are thus balanced by
the decrease in the stellar binding energies.   
This statement of energy conservation reads,
\begin{equation}
\label{eq:tidal-energy-balance}
M_* \psi_{\rm T} + M_1 \psi_* = \frac{\beta GM_*^2}{R_*} -  \frac{\beta GM_*^2}{R},
\end{equation}
where $\psi_{\rm T}$ and $\psi_*$ are evaluated along the line of centres ($\theta =0$),
for a separation $r=a$ and a stellar radius $R$ (instead of $R_*$).
The structural constant in the self gravity term (right-hand site)
is $\beta =3/(5-n)$ for a polytrope\footnote{In Sec. 
\ref{polytropicstar} we will discuss polytropic models in 
greater detail.} of index $n$ (cfr. Lai \& Shapiro 1995).

\begin{figure*}
\begin{center}
\includegraphics[width=\columnwidth]{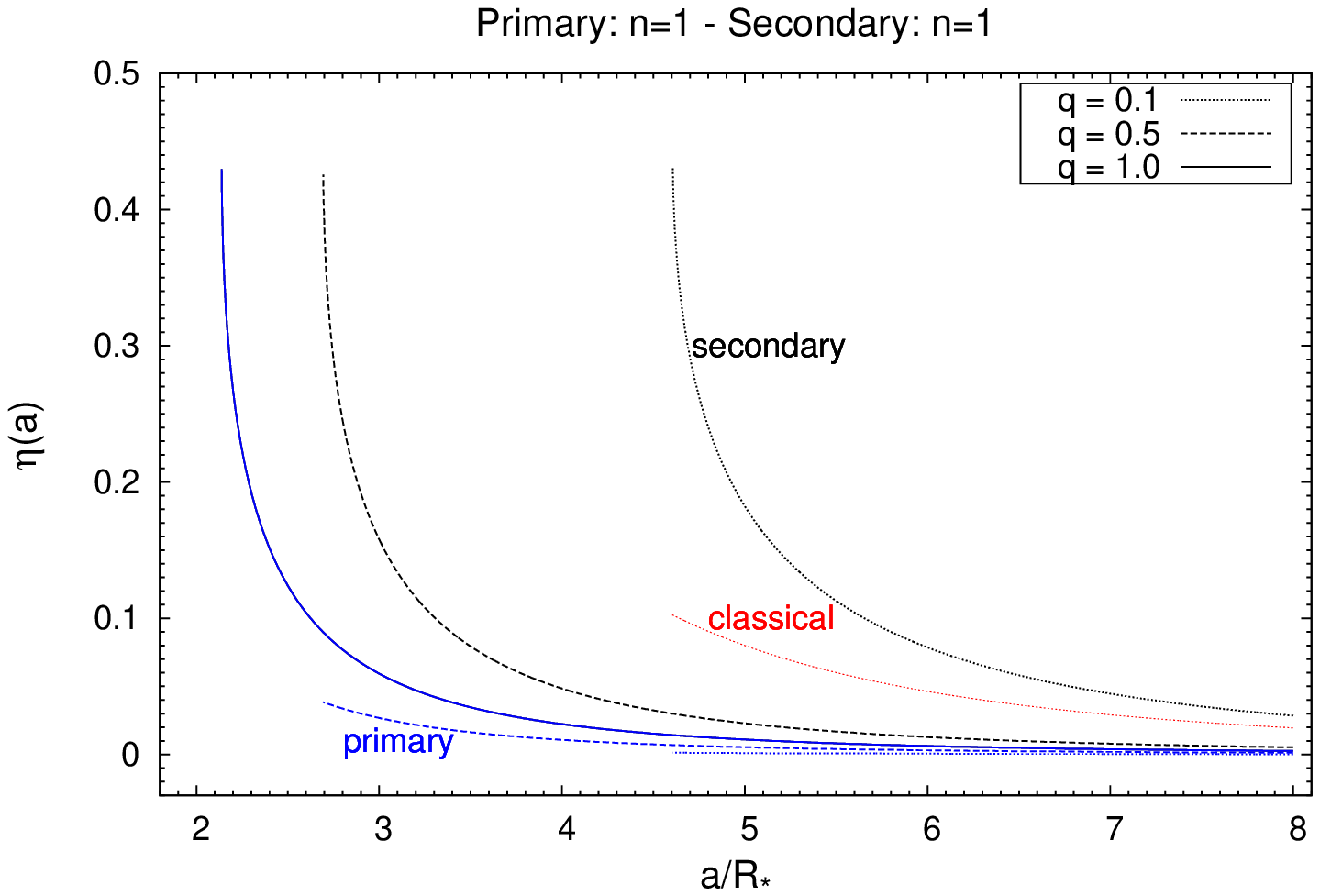}
\includegraphics[width=\columnwidth]{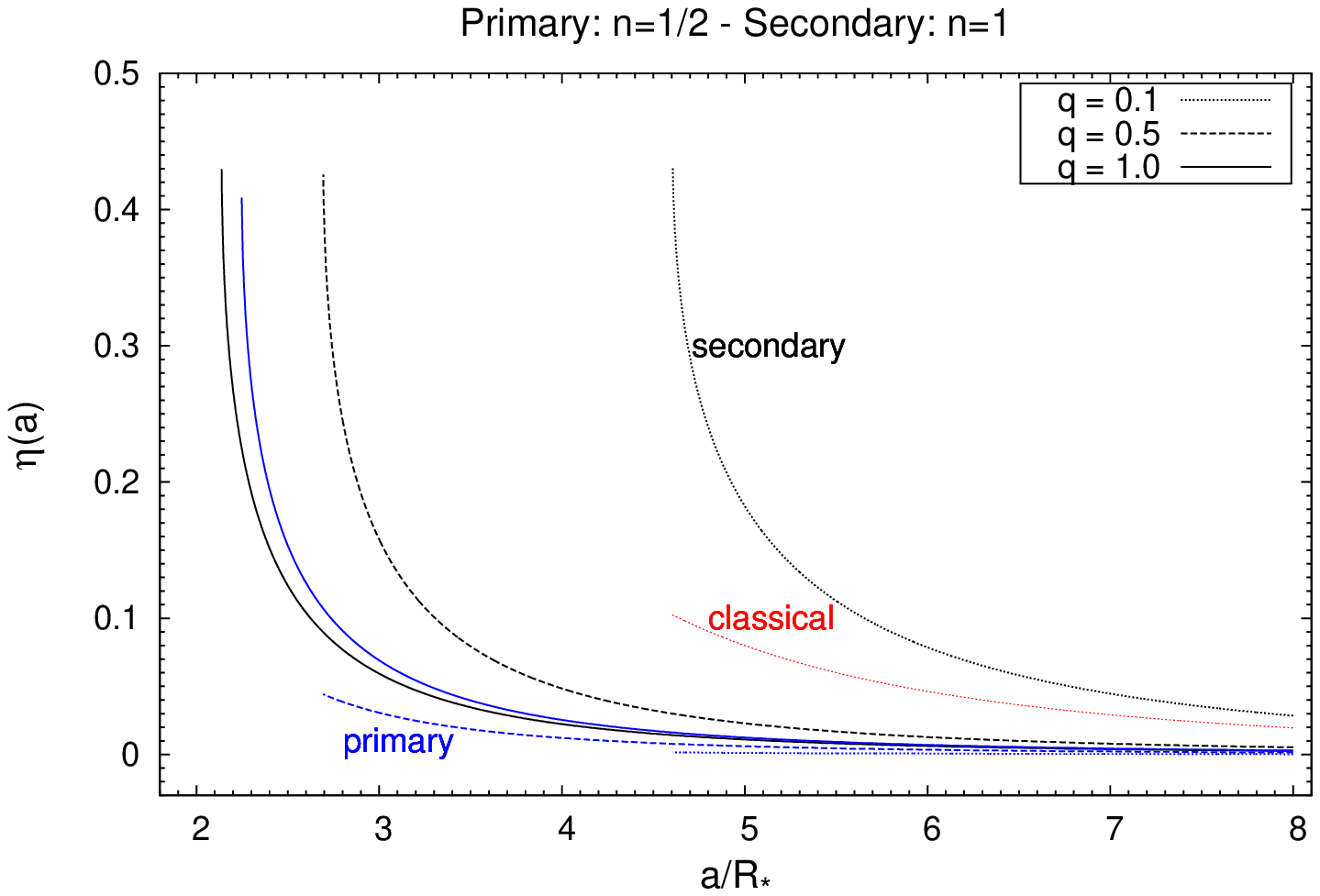}
\caption{Fractional tidal elongation, $\eta \approx h/R_*$, of the radii of both 
components in a NS-NS system, as a function of the orbital
separation. \textit{Left Panel:} two NSs described by the same e.o.s., for three
different values of their mass ratio. The black lines are for the secondary, the blue ones for the primary star.
Note that for $q=1$ the two curves are
exactly superimposed, since the NS are identical. \textit{Right Panel:} the
secondary NS is described by an $n=1$ polytrope and the
primary one by a stiffer, $n=1/2$ eos. Note that, for $q
=1$, the stiffer NS has a larger deformation, at a given $a$, and is
disrupted at a slighlty larger orbital separation.  
This is because it is less centrally condensed, and its Love number is
thus larger. On the other hand, for 
$q \leq 0.95$ the less massive star is more deformed and tidaly disrupted, dispate the steeper density profile.}
\label{fig:eta}
\end{center}
\end{figure*}

A manipulation of eq. \ref{eq:tidal-energy-balance}
leads to the following algebraic equation for $\eta$,
\begin{equation}
\label{eq:define-eta}
\kappa_2 (1+\eta)^6 + q \left(\frac{a}{R_*}\right)^3 (1+\eta)^3 - q^2 \beta
\left(\frac{a}{R_*}\right)^6 \eta = 0 \, ,
\end{equation}
For a given polytropic index $n$ and the mass ratio $q$, we can solve
numerically the above equation to obtain $\eta(a)$. In Fig. \ref{fig:eta} we show 
our results for a NS-NS system, adopting different values of 
$q$ and $n$. In the left panel, we show a binary system where both NSs are described by an 
$n=1$ polytropic equation of state (eos) , a good approximation for fiducial NSs with M=$1.4$ 
M$_{\odot}$ and R=12 Km (cfr. Sec. \ref{polytropicstar}). In the right panel, we consider 
a primary NS which is significantly more massive, $\approx 2$
M$_{\odot}$. This case can be represented by a stiffer eos with $n=1/2$ polytrope\footnote{In a BH-NS system only the secondary,
  i.e. the NS, would be deformed. This is just a sub-case of the ones 
plotted in Fig. \ref{fig:eta}.}.
%%%%%%%%%%%%%%%%%%%%%%%%%%%%%%%%%%%%%%%%%%%%%%%
%%%%%%%%%%%%%%%%%%%%%%%%%%%%%%%%%%%%%%%%%%%%%%%%%%%%%%%%%%%%%

From the above results, we obtain an unambiguous definition of the tidal 
radius as the orbital separation below which the residual binding energy of the 
secondary is unable to balance the tidal field. This always occurs when $\eta$ 
reaches a maximum  value of $\eta_{\rm T} \approx 0.4$, independently of $q$ 
and barely sensitive to $n$ (see Fig. 2). This deformation corresponds to a ratio
of semi-minor, $R_*(1-\eta/2)$, to semi-major, $R_*(1+\eta)$, axis of $0.55$.
Indeed, no real solution to eq. \ref{eq:define-eta} exists below the minimal orbital separation $a_{\rm T}$
corresponding to $\eta_{\rm T}$,

\begin{equation}
\label{eq:tidalradius}
a_{\rm T} = (2.14 \div 2.30) \frac{R_*}{q^{1/3}} \,,
\end{equation}
where the numerical coefficient depends {\em only} on $n$,
with the smallest coefficient obtained for $n=1$ and the largest one for a
$n=1/2$.  We can thus confirm the classical scaling, $a_{\rm T} \propto q^{-1/3}$.
Note that this expression for the tidal radius holds 
in general, whether the primary is a NS or a BH. 

Given the simplicity of our treatment, eq. \ref{eq:tidalradius} compares
remarkably well to recently published results of relativistic numerical
simulations (Taniguchi et al. 2008, Pannarale et al. 2011). 
For example, Pannarale et al. (2011) find that the tidal radius in their simulations
corresponds to a stellar axises ratio of $0.44$, which is not so different from our newtonian 
result ($0.55$). Indeed, our definitions (eq. \ref{eq:tidal-energy-balance} and \ref{eq:define-eta}) 
provide a simple physical argument for interpreting published numerical results.

Finally, we conclude this section by comparing our results for $\eta$ with the classical approximation
$h/R_* =  \psi_{\rm  T}/(\beta GM_*/R_*)$, 
where $\phi_*$ is not included  (solid red line in Fig.2). This approximation clearly fails to describe the final stages of the
inspiral and to determine the value of the tidal radius.

\subsection{NS-NS binaries}
\label{subsec:NSNS}

We can now describe the final stages of coalescence of two NSs,
tracking the growth in radius of both components.
For convenience, we will label
$R_1$ and $R_2$ the perturbed stellar radius of the primary and secondary star, respectively.

\begin{figure}
\includegraphics[width=0.47\textwidth]{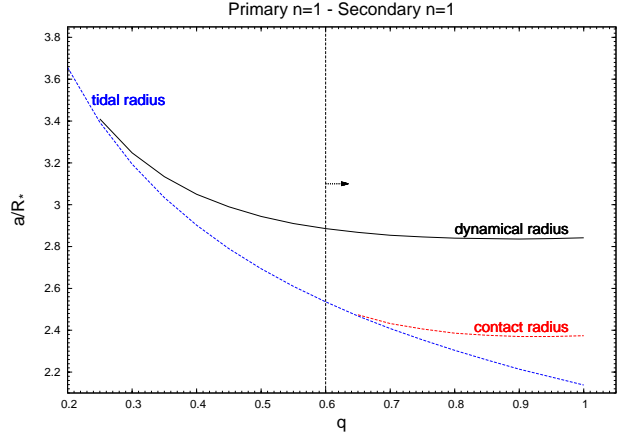}
\caption{Tidal radius ($a_{\rm T}$), contact radius ($a_{\rm c}$) and radius
  of dynamical instability ($a_{\rm d}$) as a function of the mass ratio
  $q$, for a NS-NS binary where both  components are described by an $n=1$ 
polytropic eos. Dynamical instability always sets in before 
contact. The specific eos was chosen for the sake of illustration, but
our conclusions have a general validity for any eos.}
\label{fig:dyn}
\end{figure}

Beside the tidal radius (eq. \ref{eq:tidalradius}), there is another
characteristic radius at which the inspiral may end: 
the orbital separation $a_{\rm c}$ at which the NSs first come into contact.
The question is which one is the larger.
In Fig. \ref{fig:dyn} we show $a_{\rm T}$ and $a_{\rm c}$
as a function of $q$ for a specific case of a binary with identical NSs.
For mass ratio $q \gtrsim 0.7$, $a_{\rm c} \gtrsim a_{\rm T}$.
In fact, this is a quite general result.
At the tidal radius $\eta \approx 0.4$ for the secondary star, implying that 
its perturbed radius is $R_2 \approx 1.4 R_*$. The primary's radius 
will also be deformed, by an amount which is $\sim q^2$ times smaller than 
the secondary's. The sum of the two radii will thus be $R_2(a_{\rm T}) +
R_1(a_{\rm T}) \approx (2.4+0.4 q^2) R_*$, which for $q \gtrsim 0.7$ 
is larger than the tidal radius for any polytropic index. 
For smaller values of $q$, the 
tidal radius would instead be reached before contact. However, mass ratios 
smaller than $0.7$ are highly unlikey for NS binaries, and we do not consider
them here. We thus conclude that physical collision will generally occur prior to
tidal disruption in NS-NS systems.

The last characteristic separation to compute is $a_{\rm d}$,
where the system becomes dynamically unstable.
The total 
effective potential for two NSs is

\begin{eqnarray}
\label{eq:phitot}
\psi_{\rm tot} & = & \frac{L^2 (1+q)}{2 M_* a^2}- \frac{M_* M_1}{a} -  \frac{M_*
  M_1}{a^3} (R^2_2+R^2_1) \nonumber \\
 & - & \frac{\kappa^{(2)}_2  M^2_1 R^5_2}{a^6} - \frac{\kappa^{(1)}_2
  M^2_* R^5_1}{a^6} \,,
\end{eqnarray}
Here the first term is the centrifugal potential, the only 
repulsive term in the sum, where $L^2 = L_2^2+L_1^2$ is the total angular momentum. The second represents the interaction potential 
between point masses. The third and forth are the two leading-order terms in 
their mutual tidal interaction ($\psi_{\rm  T}$ and $\psi_*$),
evaluated along the line of centres ($\theta =0$).
In the above, the perturbed stellar radii $R_{1,2}$ are a function of $a$ (see previous section), 
but, we omitted to write the $a$-dependence for simplicity.

The above potential will have a minimum - corresponding to a stable, circular
orbit - if and only if there exists a point where its first radial derivative 
is zero and its second derivative is positive. In the opposite case, there
will be no minimum and the system will be dynamically unstable 
(cfr. Clark \& Eardley 1977). Formally, derivatives of $R$ with
respect to $a$ should also be taken to derive the instability limit in this
way. Such a rigorous approach would greatly increase the difficulty of the 
calculation, adding little physical value. In fact, it is possible to show that
for $\eta \ll $ a few tenth, the terms with $dR/da$ can indeed be neglected.
\footnote{Let's take the
  derivative with respect to $a$ of any of the terms in $\sim R^2/a^3$ in eq. \ref{eq:phitot}. We then impose 
$2 (R/a^3) |dR/da|  \ll (3R^2/a^4)$ or equivalently~$\left| {\rm d}\eta/ {\rm d}a\right| \ll (3/2)
  (1+\eta)/a$. For a power-law dependence of $\eta \propto
(R_*/a)^{\delta}$, this conditions becomes $\eta \ll 3/(2\delta -3)$.
For any reasonable $\delta$ ($\delta < 10)$ this implies $\eta \ll $ a few tenth. A consistent result is obtained considering the terms in $R^5/a^6$.}
This approximation will lead us to find that the dynamical instability sets in for $\eta \le 0.1$,
justifying {\textit{a posteriori}} our procedure.

With our approximation well defined, we now impose the first derivative of  eq. \ref{eq:phitot} to be zero,

\begin{eqnarray}
\label{eq:first-deriv-zero}
\frac{L^2(1+q)}{M_*a^3} & = & \frac{M_* M_1}{a^2}+ \frac{6 \kappa^{(2)}_2 M_1^2
  R^5_2}{a^7} \nonumber \\
 & + &  \frac{6 \kappa^{(1)}_2 M_*^2
  R^5_1}{a^7} + \frac{3 M_1 M_*}{a^4} (R^2_1+R^2_2),
\end{eqnarray}
while requiring the second derivative to be positive,
\begin{eqnarray}
\label{eq:second-deriv-positive}
\frac{3 L^2(1+q)}{M_* a^4} & \geq & 2 \frac{M_1 M_*}{a^3} + 
\frac{42 \kappa^{(2)}_2  M_1^2
  R^5_2}{a^8} \nonumber \\
 & + & \frac{42 \kappa^{(1)}_2 M_*^2  R^5_1}{a^8} + \frac{12 M_1 
M_*}{a^5} (R^2_1+R^2_2)\, .
\end{eqnarray}
Combining the two conditions above and after some manipulation, we obtain
an equation for the adimensional radial distance, $y_{\rm d}=a_{\rm d}/R_2(a_{\rm d}) $, where the instability sets in: 

\begin{equation}
\label{eq:dynamical}
y_{\rm d}^5 - 3 y_{\rm d}^3 \left[1+\left(\frac{R_1}{R_2}\right)^2\right]
\geq \frac{24~\kappa^{(2)}_2}{q} \left[1+q^2
  \frac{\kappa^{(1)}_2}{\kappa^{(2)}_2} \left(\frac{R_1}{R_2}\right)^5 
\right].
\end{equation}

To test the validity of our approximation, we considered 
the specific case studied by Lai 
\& Shapiro (1995) of a NS-NS binary, where the stars have the same 
radius $R_{\rm 0} =12$ km, polytropic index
 $n=1/2$ and have a mass ratio $q=1/2$.
These authors solve numerically the full set of (newtonian) hydrodynamical equations and
find that the instability ensues when 
$a_{\rm d}/R_* \simeq 3.1$ (cfr. top panel of their Fig. 2). Our analytical method
(eq. \ref{eq:dynamical}) gives $a_{\rm d}/R_* \simeq
 3.05$. Considering the significantly different procedures, the
 agreement is quite remarkable.

We can now proceed to solve equation \ref{eq:dynamical} in other cases of interest.
In Fig.\ref{fig:dyn}, we plot the solution as a function of the mass ratio, 
when the secondary is described by an $n=1$ polytropic  eos.
It shows that the dynamical instability always occurs {\it before} the two NSs come into 
contact (cfr. Lai, Rasio \& Shapiro 1994a; Lai \& Shapiro 1995).

%%%%%%%%%%%%%%%%%%%%%%%
%%%%%%%%%%%%%%%%%%%%%%%

\subsection{BH-NS binary}
\label{subsec:BHNS}

Before repeating the same calculation for a BH-NS binary, we 
briefly address the notion of a ``contact'' radius in such systems.
We consider the event horizon ($R_{\rm h} = 2GM_1/c^2$) as the relevant 
radius for a (non-spinning) BH, and assume that the secondary will 
disappear in  the event horizon once its elongated radius first touches it. 
Thus, the contact radius $a_{\rm c}$ in a BH-NS binary is the 
orbital separation at which the NS is swallowed in the event horizon.

We now ask what is the condition for the tidal radius to be 
reached before contact, so that tidal disruption will occur in an interesting 
region. The sum of the two radii when the tidal radius $a_{\rm
  T}$ is reached can be written as
\begin{equation}
R_2(a_{\rm T})+R_{\rm h} \approx  2 R_*
\left(0.7+\frac{{\cal{C}}}{q}\right) \, ,
\end{equation}
where we defined the NS compactness ${\cal{C}} = GM_*/(c^2 R_*)$ and used the
fact that $\eta_{\rm T} \approx 0.4$. 
Using eq. (\ref{eq:tidalradius}), and after some manipulation, the condition 
$a_{\rm T} > a_{\rm c}$ leads to the inequality,

\begin{equation}
\label{thirdorder}
q - (1.53 \div 1.64) q^{2/3} +\frac{{\cal C}}{0.7} \leq 0 \, .
\end{equation}

The above condition can be fulfilled for $q$-values that depend on
the NS compactness. For our fiducial NS, the relevant coefficient 
is the smaller one. This gives $q \gtrsim 0.11$, or M$_1 \lesssim 12.7$
M$_{\odot}$. On the other hand, the 
fiducial NS will disappear in the BH horizon before being disrupted if the BH 
is more massive than $\approx 12.7$ M$_{\odot}$. 
Changing the NS structure gives slightly different numerical values,
 without affecting our  conclusion that for stellar mass black holes,
 the tidal disruption generally occurs outside the event horizon.

\begin{figure*}
\begin{center}
\includegraphics[width=\columnwidth]{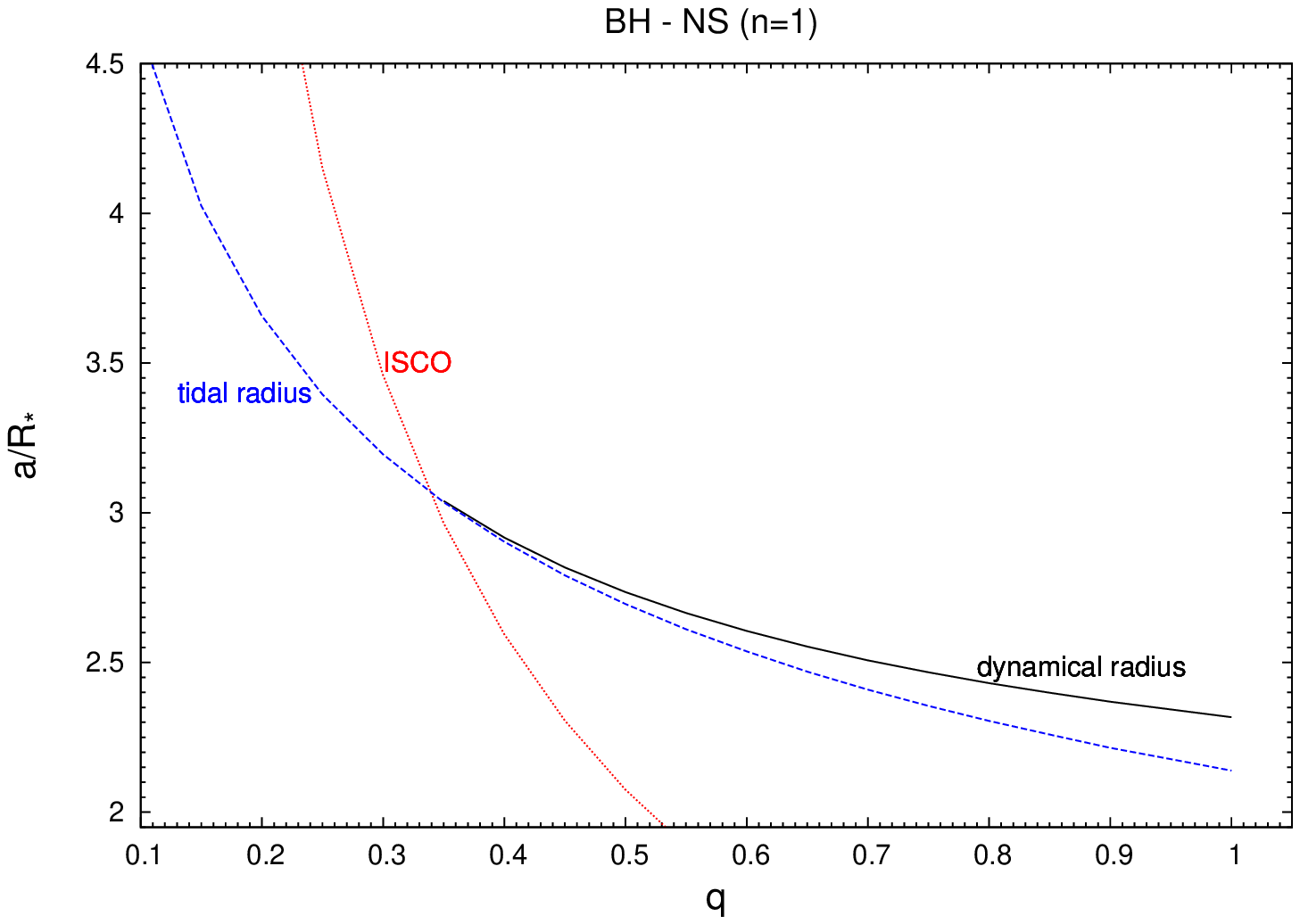}
\includegraphics[width=\columnwidth]{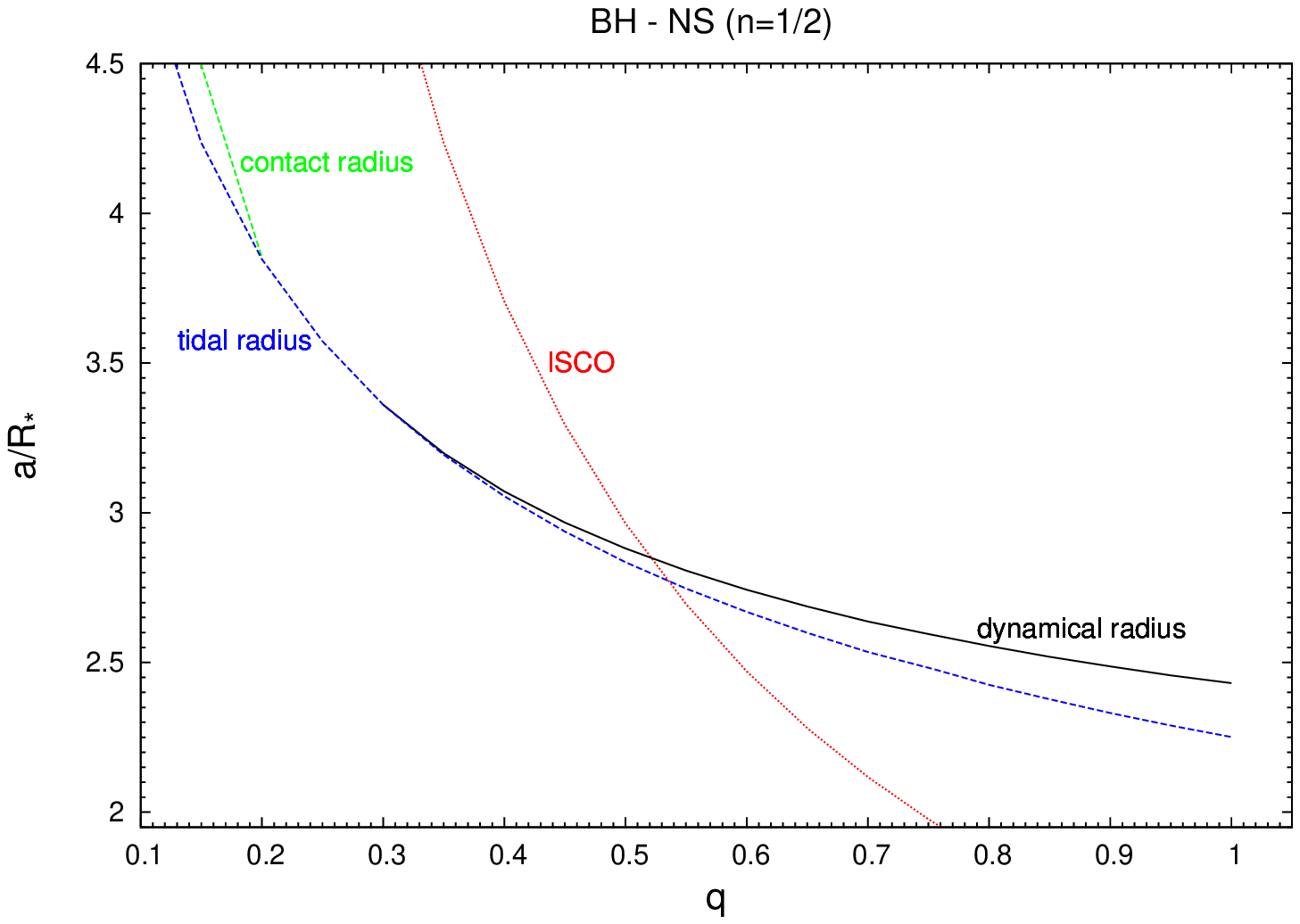}
\caption{Tidal radius ($a_{\rm T}$), contact radius ($a_{\rm c}$), radius
  of dynamical instability ($a_{\rm d}$) and ISCO as a function of the mass ratio
  $q$, for BH-NS binaries. \textit{Left Panel:} the secondary NS  has $M_*=1.4$ M$_{\odot}$, R$_* = 12$ km and
  $n=1$. \textit{Right Panel:} the secondary NS  is a massive, 2 M$_{\odot}$
  NS with R$_*$ = 12 Km and an $n=1/2$ eos.}
\label{fig:raggiBH}
\end{center}
\end{figure*}
 A well-known property of the Schwarzschild potential is its 
steepening at close orbital separations. This genuinely relativistic effect
determines the so-called innermost stable circular orbit (ISCO). Although 
different in nature, this is totally equivalent to the dynamical instability 
discussed before, for our purposes. Indeed, once this separation is reached, 
orbits cannot be closed anymore and the secondary will plunge 
into the event horizon on a dynamical timescale. Therefore, we also want to establish the location of
the ISCO with respect to the tidal radius, in order to determine its role in 
the orbital evolution of a BH-NS binary.
For a Schwarzschild BH, $a_{\rm 
isco} = 3 R_{\rm h}$.  Using
eq. \ref{eq:tidalradius} for the tidal radius it is straightforward to see
that the ISCO will be {\em larger} than 
the tidal radius if $q \lesssim (0.3 \div 0.33)
\left({\cal{C}}/0.17 \right)^{3/2}$, for $1/2 \leq n \leq 1$.

Finally, we discuss the condition for dynamically
  instability in BH-NS systems. The reasoning is identical to that in the previous section
  but the formulae must be slightly modified
to account for the fact that the BH does not suffer
tidal deformations (its Love number is zero). As a consequence, we expect the 
dynamical instability to set in at smaller orbital separations, for a given 
set of system parameters. It is a simple matter to re-write 
eq. \ref{eq:dynamical} for a BH-NS system and obtain,

\begin{equation}
\label{eq:dynamical-BH}
y_{\rm d}^5 - 3 y_{\rm d}^3 = 24~\frac{\kappa_2}{q},
\end{equation}

The solution of this equation as a function of $q$ is plotted in Fig. 
\ref{fig:raggiBH}, along with the tidal radius, the contact radius
and the ISCO, for two different BH companions.
The main result is that for a BH with masses greater than a few $M_{\sun}$,
the relativistic dynamical instability sets in {\em before} the tidal one (i.e. $a_{\rm isco} > a_{\rm d}$).
In this regime, the secondary is generally tidally disrupted --as opposed to swallowed --
for stellar mass black holes.

\section{Summary of the evolutionary scenarios}

\label{sec:summary-scenarios}
In the previous sections we analyzed several physical mechanisms that produce 
important non-trivial effects in the late evolution of coalescing
binaries.These become important at different orbital separations.
In the following, we summarize our main results for the ordering of these
phases, in NS-NS and NS-BH systems.

\subsection{NS-NS binaries}
The situation is relatively simple in these systems. The orbit becomes
dynamically unstable at the orbital separation $a_{\rm d}$, which is larger
than any other relevant size. Subsequently, the two NSs will inevitably touch 
before reaching the tidal radius (eq. \ref{eq:tidalradius}). This gives the 
following ordering: $a_{\rm T} <  a_{\rm c} < a_{\rm d}$.

The orbital evolution of a NS-NS system will thus be driven by the emission of 
GWs down to $a_{\rm d}$. Once in the dynamically unstable region, the binary orbital shrinkage
is significantly accelerated and occurs on the dynamical timescale 
($t_{\rm d}$) at $a_{\rm d}$. The NSs eventually touch at $a_{\rm
  c}$. The inspiral ends here and the complex merger phase sets in.
Correspondingly, we expect the tidal torque to be characterized by a lag angle 
$\alpha_{\rm  GW}$ during the first part, and by a larger lag 
$\alpha_{\rm d} \sim \alpha_{\rm GW} (t_{\rm GW}/t_{\rm d})$ once the
dynamical instability sets in. This will be discussed in greater detail in the
next sections.

\subsection{BH-NS binaries}

The mass ratio in these binaries turns out to have a large effect in 
determining the relative locations of the characteristic radii. The dynamical 
instability occurs in these systems only for relatively low mass BHs. However, 
we noticed that the ISCO can play a similar role at significantly larger 
orbital separations, when the BH mass grows. The contact radius is generally 
smaller than the tidal radius, and only for very small mass
ratios ($\lesssim 0.1$) it plays a role. We neglect such small $q$ values here. 
 As a consequence, we expect three types of orbital evolution for BH-NS 
binaries, depending on the mass ratio. For a fiducial NS mass of 1.4 
M$_{\odot}$ and radius f 12 Km, this will be a function of the BH mass only. 

\begin{itemize}

\item if M$_{\rm 1} \lesssim 4$ M$_{\odot}$ we have the ordering $a_{\rm
  ISCO} < a_{\rm T} < a_{\rm d}$.
 In this case, GWs drive orbital shrinking until the dynamical instability is
 reached. A short plunging phase follows, until the NS reaches the slightly
smaller tidal radius at which it gets tidally disrupted.
The orbital evolution stops at $a_{\rm T}$;

\item if M$_{\rm 1}> 4$ M$_{\odot}$ the ISCO will correspond to the largest 
orbital separation. In this case, GWs drive the orbital evolution until the NS 
reaches the ISCO. A plunging phase follows also in this case, whose radial
extent grows quickly with increasing M$_{\rm 1}$. The orbital
evolution will again be stopped at $a_{\rm T}$;

\item if M$_{\rm 1} \gtrsim $14 M$_{\odot}$ we get 
$a_{\rm d} < a_{\rm T} < a_{\rm c} < a_{\rm isco}$. Also in this case, 
GW-driven orbital shrinking will be overtaken at $a_{\rm isco}$ by a plunging
phase. The NS is swallowed in the horizon before being tidally disrupted. 
\end{itemize}

\section{The total tidal energy}
\label{sec:totaltidalenergy}
The bulge has angular momentum but also a rotational energy given by 
$E_{\rm b} = (1/2) I_{\rm b} \omega_{\rm b}^2$.
If not dissipated, this energy remains stored in the wave and increases as the
orbit shrinks.
We now evaluate the total amount of energy $E_{\rm b,T}$ that the tidal torque
transfers to the secondary's bulge, as the system evolves towards coalescence. 
Since the lag angles considered are small, we approximate  $\omega_{\rm b}
\simeq \omega_{\rm o}$ in order to simplify the calculations. This implies that 
the gain of energy is $dE_{\rm b} =  3 I_{\rm b} \omega_{\rm o} d\omega_{\rm o}$,
regardless of the speed of the inspiral.
Thus the total energy released up to an orbital frequency $\omega_{\rm o}$ is

\begin{eqnarray}
\label{eq:DE_tot}
 E_{\rm b,T}(\omega_{\rm o}) &= & \frac{18 \pi^2 \kappa_2 M_* R^2_*}{(1+q)^2 \omega^4_*} \int_{0}^{\omega_{\rm o}} \omega_{\rm o}^5 d\omega_{\rm o} \nonumber \\
 &= & \frac{3 \pi^2 \kappa_2 M_* R^2_*}{(1+q)^2 \omega^4_*}  ~\omega_{\rm o}^6  , 
 \end{eqnarray}
 where we assumed an infinite initial orbital separation. This is justified by the fact that essentially all of the energy
is transferred close to the final radius, and thus the starting point of the integration is irrelevant. 
 In the most interesting cases, the inspiral ends at the tidal radius, where the orbital frequency  is  $\omega(a_{\rm T}) \equiv \omega_{\rm T}$,
 
 \begin{eqnarray}
\label{tidalradius-two} 
\omega_{\rm T} =  c_1~\omega_* (1+q)^{1/2}  \approx  (2.7 \div 3) \times 10^3  \left(\frac{M_*}{1.4~{\rm M}_{\odot}}\right)^{1/2} \\ \nonumber \left(\frac{12~{\rm Km}}{R_*}\right)^{3/2} (1+q)^{1/2} 
 {\rm \frac{rad}{s}}\, .
\end{eqnarray}
 In the above, the coefficient $c_1 = (0.29 \div 0.32)$, where the
  smaller value is for $n=1/2$ and the larger is for $n=1$. Thus, from eq. \ref{eq:DE_tot}  
 \be
  E_{\rm b,T} \approx 3 \times  10^{51} \frac{E_*}{4.3 \times 10^{53}} \frac{\kappa_2}{0.26} (1+q) \left(\frac{\omega_{\rm o}}{\omega_{\rm T}}\right)^6 ~{\rm erg}, \nonumber
\label{eq:DE_tot_tidal}
 \ee
 where $E_* = GM_*^2/R_*$ is the secondary binding energy. The Love number $\kappa_2$ was normalized to a 
typical value for an $n=1$ newtonian polytrope\footnote{Relativistic models
  give $0.05 \lesssim \kappa_2 \lesssim 0.1$ (Hinderer 2008; Postniakov
  2009). We adopt the newtonian value to compare with our results of \S \ref{polytropicstar}.} and 
the orbital frequency at the tidal radius is given by
eq. \ref{tidalradius-two} with $c_1=0.32$ for an $n=1$ polytropic NS.

The total energy transferred to the bulge in an equal mass binary is thus $ \sim 10^{-3}-10^{-2}$ 
of the secondary binding energy. 
The numerical coefficient in the last line of eq. \ref{eq:DE_tot_tidal} is for our fiducial NS parameters. 
A solar mass WD, with $R_* = 5 \times 10^3$ km, has instead
$E_* \approx 5 \times 10^{50}$ erg and $E_{\rm b,T} \approx  1.3 \times 10^{48}$ erg
(for WDs $\kappa_2 \sim 0.1$, cfr. Verbunt \& Hut 1983).
 
 The orbital energy  $E_{\rm b,T}(\omega_{\rm T})$ which is transferred  to the secondary
 can be of the same order of that required for co-rotation at  $a_{\rm t}$: $\simeq (1/2) (0.35) M_*
 R_*^2 \omega_{\rm T}^2 \approx  0.02 E_*$. However,   $E_{\rm b,T}$ is mostly stored in the quadrupolar
  oscillations within the secondary and only a {\em minor} fraction ($t_{\rm
  GW}/t_{\rm v}$) is actually viscously converted into spin energy.
We, therefore,  confirm that binaries with compact objects cannot reach co-rotation before they merge 
(Kochanek 1992; Bildsten \& Cutler 1992).

In the following, we will show that this internal motion, i.e. the rotation of the tidal bulge,
is  a function of radius within the secondary. In this respect, it inherits the differential nature of the tidal force.

\section{The differential structure of the tidal bulge}
\label{subsec:diffrot}

Our description of the tidal bulge is that of a wave which rotates in 
the secondary's frame at the rate\footnote{Strictly speaking, it is $\simeq
\omega_{\rm o}    - \omega_{\rm s}$, but we are neglecting the secondary's
spin.} $\omega_{\rm b} \lsim \omega_{\rm o}$, exciting quadrupolar fluid
oscillations. As such, 
it carries a total angular momentum, $J_{\rm b}$ (\S~\ref{subsec:fix_separation})
and rotational energy $E_{\rm b}$ (\S~\ref{sec:totaltidalenergy}). 
We can thus associate to the wave
an ``effective moment of inertia'' $I_{\rm b}$ (eq.~\ref{eq:Ib}), such that 
$J_{\rm b}\equiv I_{\rm b} \omega_{\rm b}$ and $E_{\rm b} \equiv (1/2) I_{\rm b} \omega_{\rm b}^2$.
If circumstances arise that lead the wave to rotate slower than
the tidal potential, the system seeks immediately a new equilibrium configuration,
accelerating the wave, via a tidal torque $N_{\rm T}$.
We showed that this happens when GWs drive orbital acceleration
$\dot{\omega}_{\rm o}$, 
or when the orbit becomes dynamically 
unstable, at small orbital separations (either at $a_{\rm d}$ or $a_{\rm isco}$). 
In those cases, we argued
that mechanical equilibrium requires the tidal 
torque to accelerate the bulge at the same rate, $\dot{\omega}_{\rm b} = 
\dot{\omega}_{\rm o}$.
We now proceed to calculate the radial structure of this
rotating quadrupolar tide (or wave). To this aim, we divide it in
concentric rings and assign to each ring the mass contained in the 
corresponding spherical shell of the NS. We then require
each ring to be in mechanical equilibrium.
Given the marginal role played by viscosity in the systems of interest,
we can assume that there is no transfer of energy and momentum both among
the rings and between the tidal bulge as a whole and the stellar spin. 

The total tidal torque $N_{\rm T}$
depends on 3 global quantities of the secondary (see eq.\ref{classicaltorque}): its radius, Love number $\kappa_2$ and lag angle $\alpha$. 
Hence, the same expression holds for any radius $r<R_*$, with the appropriate global values of $\kappa_2$ and $\alpha$.
The difference between the torques applied to two concentric portions of the star, with 
radii $r$ and $r+dr$, will thus constitute the torque applied to the ring with inner radius $r$.
To first order in $dr$, we get
\begin{equation}
\label{define-difftorque}
d N_T(r) = 3 \frac{G M^2_1}{a^6} d(\kappa_2 r^5  \alpha).
\end{equation}
where we leave indicated a generic lag angle $\alpha$.

The equilibrium condition demands that this torque accelerates the ring so to track the orbital acceleration:
$\dot{\omega}_{\rm b}(r)=\dot{\omega}_{\rm o} =$ const.
The ring angular acceleration is  $\dot{\omega}_{\rm b} = dN_{\rm T}/dI_b$, where the right-hand term can be
written as the ratio between

\begin{eqnarray}
\label{define-radial-derivs}
\frac{dN_T}{dr} & = & \frac{3 \omega^4_o}{G (1+q)^2} ~ 
\frac{d}{dr}(\kappa_2 r^5 \alpha), 
\label{eq:nt_r}
\end{eqnarray}
and
\begin{eqnarray}
\frac{dI_b}{dr} & = & \frac{6 \pi^2 \omega^4_o}{G^2 (1+q)^2} 
\frac{d}{dr} \left[\frac{\kappa_2 r^8}{M(r)}\right], 
\label{eq:ib_r}
\end{eqnarray}
where the last equation is obtained by taking the derivative with respect to $r$ of eq.~\ref{eq:Ib}.
Therefore,
\begin{equation}
\label{alpha_gw_shell}
\dot{\omega}_{\rm b}=\frac{dN_{\rm T}}{dI_{\rm b}} = 2 \pi^2 G f(r) = \dot{\omega}_{\rm o},
\end{equation}
\no
where $f(r)$ is the ratio of the two radial derivatives in eq.~\ref{eq:nt_r} and eq.~\ref{eq:ib_r}.
Solving eq.~\ref{alpha_gw_shell} for the lag angle $\alpha$, we obtain
\be
\label{alpha_gw_radial}
 \alpha (r)= 2 \pi^2 \dot{\omega}_o/ \omega^2_*(r).
\ee

\begin{figure}
\includegraphics[width=0.48\textwidth]{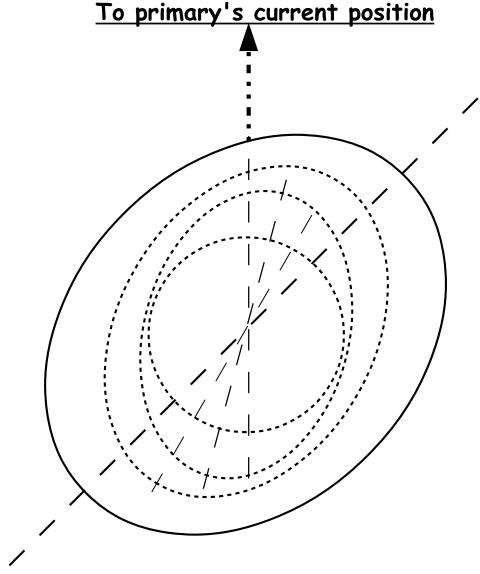}
\caption{The structure of the rotating tidal bulge, resulting from the radial
  variation of the lag angle. The crest of the wave at different radii is in
  phase with different positions of the primary along the orbit.}
\label{fig:tide_2}
\end{figure}

Generally, eq.~\ref{alpha_gw_radial} implies a tidally
distorted structure, where the lag angle 
varies within the star. This is because the restoring
force, $\omega_*(r) = GM(r)/r^3$, is usually a function of $r$,
where $M(r)$ is the mass within $r$ and $M(R_*) = M_*$.
This conclusion is sketched in Fig. \ref{fig:tide_2}.
A notable case is that of a constant density profile, which will ensure no distortion and
a lag angle given by the global value $\alpha = 2 \pi^2
  \dot{\omega}_o/\omega^2_*$ (eq.~\ref{eq:agw}).

The exact dependence of  $\alpha$ on $r$ is determined by 
the functional form of $M(r)$. In order to proceed, we thus need to consider a 
specific stellar structure.
%%%%%%%%%%%%%%
%%%%%%%%%%%%%%
%%%%%%%%%%%%%%%%%%%%%%%%%%%%%%%%%%%%%%%%%%%%%%%%%%%%%%%%%%%%%%%%%%%%%%%%%%%%%%%%%%%%
\section{Tidal effects in neutron stars with polytropic eos}
\label{polytropicstar}
In this section, we 
calculate the radial profile of a rotating quadrupolar tide, which is induced 
in a NS by the tidal field of the primary.

For this purpose, we approximate NSs as polytropic stellar models, 
where the pressure $P$ is given by $P = K_{\rm p} \rho^{1+1/n}$. Such
  a model, though idealized, describes reasonably well the NS global properties obtained
with more sophisticated equations of state (Lattimer \& Prakash 2001; Lattimer
\& Prakash 2006). 
The equation of state, together with the stellar radius and mass, completely
determine the stellar profile. In particular, a $n=1$ index is well
  suited to describe a typical NSs with a radius $R_* = 12$ Km and mass
  $M_*=1.4 M_{\odot}$. However, recent indications for a 2 M$_{\odot}$ NS
 (Lattimer \& Prakash 2010 and references therein) suggest that a NS 
eos might range in stiffness. More massive objects would thus be better 
approximated by stiffer polytropes, with $n=1/2$ or $n=2/3$. According
to the detailed analysis by Lattimer \& Prakash (2006), most eos's compatible 
with current observational constraints give radii in a fairly narrow range  
for NSs in the $(1-2)$ M$_{\odot}$ range. Therefore, we adopt the reference 
value of $R_* = 12$ Km for the radius for \textit{any} NS. On the other hand, 
we associate $n=1$ polytropes to $1.4~M_{\odot}$ NSs and either $n=1/2$ or 
$n=2/3$ to $2~M_{\odot}$ NSs.

\subsection{The differential rotation of the tidal wave}
\label{sec:diff-tidal-wave}
For a polytropic star the equation of hydrostatic equilibrium is
reduced to the Lane-Emden equation between the dimensionless radial coordinate
$\xi$ and the dimensionless density $\theta$. $\xi$ is related to
the radius by $r = A \xi$, where the constant $A^2 = (n+1) K_{\rm
  p}/(4 \pi G) \rho_{\rm c}^{(1-n)/n}$.  
The dimensionless density distribution is $\theta^n = \rho/\rho_c$, 
where $\rho_c$ is the central density. The outer boundary ($\xi_{\rm out}$) is found 
as the first zero of the solution $\theta(\xi)$ of the Lane-Emden
equation.
Knowing $\xi_{\rm out}$ and the star's physical radius, one can derive 
$A = R_*/\xi_{\rm out}$. 

\begin{figure*}
\begin{center}
\includegraphics[width=\columnwidth]{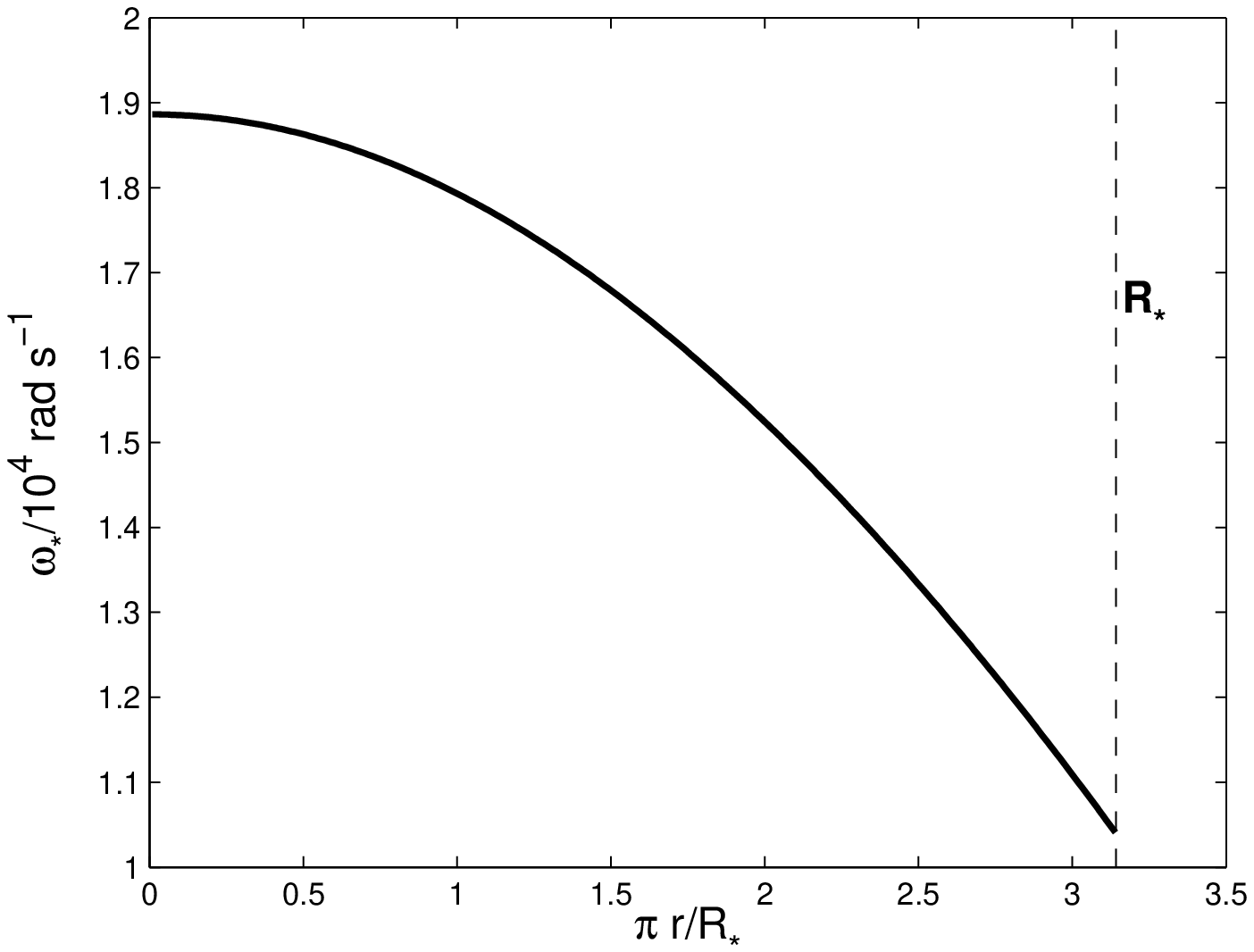}
\includegraphics[width=\columnwidth]{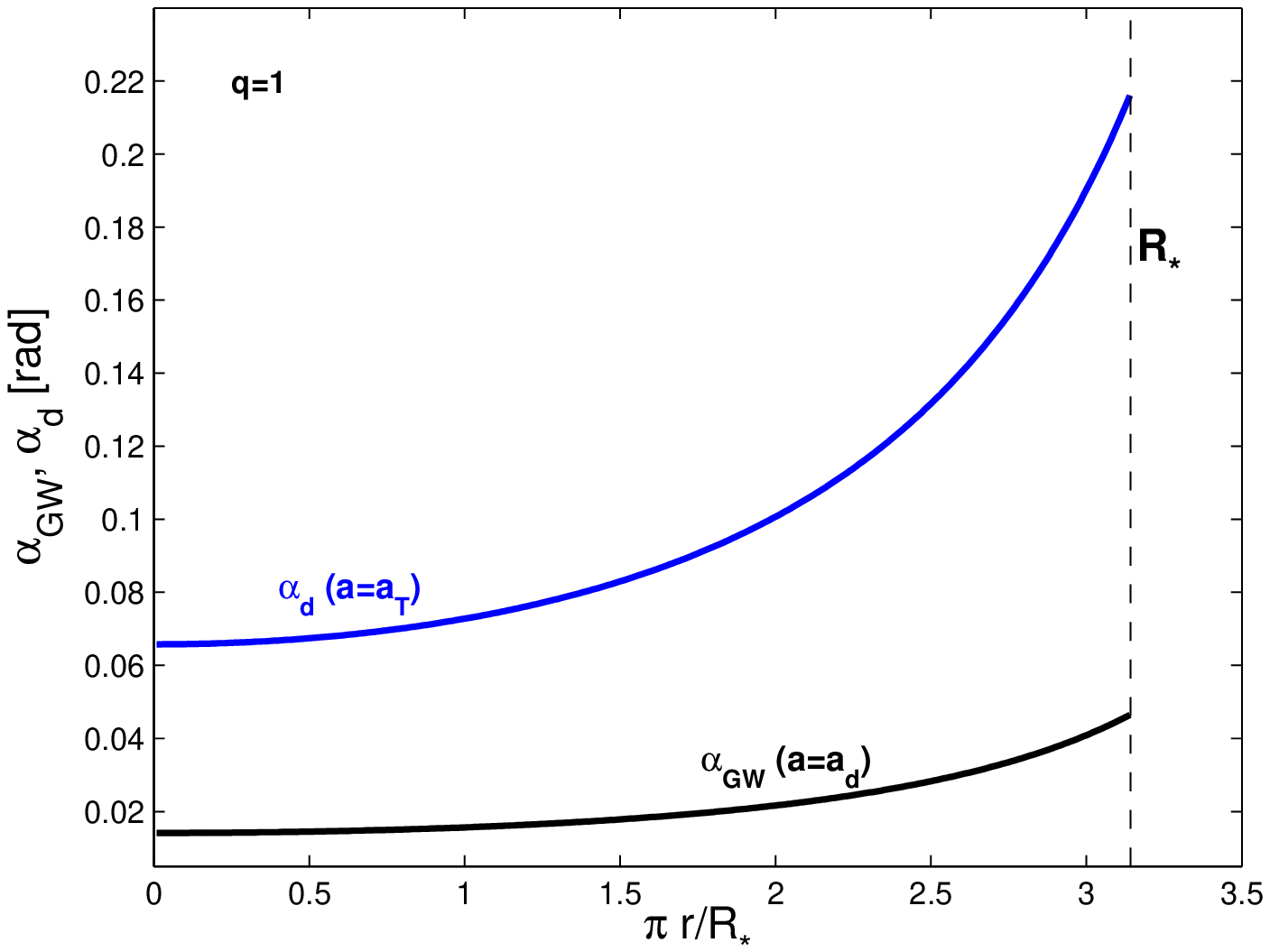}
\caption{ Radial structure for $n=1$ polytropic equation of state.
{\it Left panel: } the characteristic frequency $\omega_*$;
 {\it Right panel:}  the lag angles $\alpha_{\rm GW}$ (black line) and $\alpha_{\rm d}$ (blue line) at respectively the dynamical and tidal radius.
All quantities are plotted  as a function of the normalized radius $\xi = \pi r/R_*$. The binary mass fraction is $q=1$.}
\label{fig:tide_3}
\end{center}
\end{figure*}

Integrating $\theta$ over $\xi$ one obtains an expression for the mass
within radius $\xi$ as
\begin{equation}
\label{emmediR}
M(<\xi) = 4 \pi A^3 \rho_{\rm c} \int_0^{\xi} \xi'^2 \theta^n(\xi') {\rm
  d}\xi' =  - 4 \pi A^3 \rho_{\rm c} \xi^2 \frac{{\rm d}\theta}{{\rm
    d}\xi}.
\end{equation}
The characteristic frequency of oscillation of matter within radius $\xi$  can now be 
written as
\begin{equation}
\label{omega_char_R}
\omega^2_*(\xi) = 4 \pi G \rho_{\rm c} \frac{1}{\xi}
\frac{{\rm d}\theta}{{\rm d}\xi} = \omega^2_* \frac{\xi_{\rm out}}{\xi}
\frac{{\rm d}\theta /{\rm d}\xi}{\left[{\rm d}\theta /{\rm d}\xi\right]_{\xi_{\rm out}}}.
\end{equation}
From this equation,  we can obtain the corresponding value of the 
lag angle $\alpha_{\rm GW}$ when GWs drive the orbital evolution,
\begin{equation}
\label{alfa_gw_R} 
\alpha_{\rm GW} (\xi) = 2 \pi^2 \frac{\dot{\omega}_o}{\omega^2_*} = 2 \pi^2
\frac{\dot{\omega}_o}{\omega^2_*} \frac{F_n(\xi_{\rm out})}{F_n(\xi)}.
\end{equation}
where $F_n(\xi) \equiv (1/\xi) {\rm d}\theta/{\rm  d}\xi$.
The run of $\omega_*(r)$ from the centre to the outer boundary is shown in 
Fig. \ref{fig:tide_3} left panel. The corresponding run of $\alpha_{\rm GW}
(r)$ at $a=a_{\rm d}$  is shown in
Fig. \ref{fig:tide_3} right panel. The external layers of the bulge 
react more slowly to the change in position of the primary and lag behind
the NS core, which instead, is promptly realigned. 

With an identical reasoning, we can define a generic dynamical lag
$\alpha_{\rm d}$, that holds when the orbit evolves on a dynamical timescale
  $t_{\rm d}$,
\begin{equation}
\label{alpha_dyn_R}
\alpha_{\rm d} (\xi) = 2\pi^2 \frac{\omega_{\rm o}}{\omega_*^2 (\xi)
{\rm t}_{\rm d}} \,. 
\end{equation}
 In Fig.  \ref{fig:tide_3} right panel, we plot $\alpha_{\rm d}$ at $a = a_{\rm T}$ (blue line). The absolute
value and especially its derivative are larger than those of $\alpha_{\rm GW}$ at $a_{\rm d}$ (black line):
during the dynamical plunging the star becomes significantly more differentially perturbed.
%%%%%%%%%%%%%%%%%%%%%%%%%%%%
%%%%%%%%%%%%%%%%%%%%%%%%%%%

\begin{figure}
%%%%%%%%%%%
\includegraphics[width=0.48\textwidth]{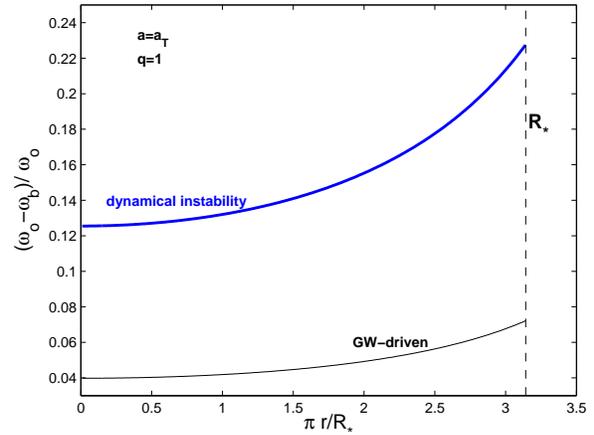}
\caption{As Figure \ref{fig:tide_3} right panel, but for the angular frequency of the wave, $\omega_{\rm b}$.
In the $y-axis$ we show the fractional distance of $\omega_{\rm b}$ to the orbital frequency $\omega_{\rm o}$.}
\label{fig:tide_4}
\end{figure}

From the above expressions, we can define the rotation frequency of each 
portion of the tidal bulge, in analogy with the classical description of the 
equilibrium tide in viscous bodies (eq.\ref{eq:alpha_v_tlag}). The lag time 
with which the bulge reacts is now $t_*(r)$, while $\omega_{\rm b}(r)$ plays
the role of the spin frequency $\omega_{\rm s}$. At a given radius $r$, the
corresponding relation between $\alpha_{\rm GW}$ (or $\alpha_{\rm
    d}$) and $\omega_{\rm b}$ is thus
\begin{eqnarray}
\label{diff-velocity}
\left[\omega_{\rm o} - \omega_{\rm b}\right] t_*(r) & = & \alpha_{\rm GW}(r) ~~~~~{\rm
  if~~~} a \geq a_{\rm d} \nonumber\\
\left[\omega_{\rm o} - \omega_{\rm b}\right] t_*(r) & = & \alpha_{\rm d}(r) ~~~~~{\rm
  if~~~}  a_{\rm T},a_{\rm c} \leq a < a_{\rm d} \, .
\end{eqnarray}
 When the orbit becomes dynamically unstable, either at $a_{\rm
    d}$ or at $a_{\rm isco}$, the system evolution will depend on the details of
  the plunging phase, whose treatment is beyond our scope. Here, we simply
  assume that, at the onset of dynamical instability, the relevant time 
will switch from the GW-driven inspiral time (eq. \ref{taugw}) to the much 
shorter dynamical time calculated at $a_{\rm d}$ : $t_{\rm d} \approx 2 \pi / \omega_{\rm d}$, where $\omega_{\rm d} \equiv \omega_{\rm o} 
(a_{\rm d})$. Once this occurs, our most 
conservative assumption will be to
freeze the further growth of the tidal lag, fixing $\alpha_{\rm d}$ to the 
value it has \textit{when the dynamical instability first  sets in}. With this
assumption, the radial behaviour of $\omega_{\rm b}$ along the orbital
evolution is completely defined. Inserting eqs. \ref{omega_char_R}, 
\ref{alfa_gw_R} and \ref{alpha_dyn_R} into eq. \ref{diff-velocity}, gives
\begin{eqnarray}
\label{omega_b_r}
\omega_{\rm b}(\xi) & = & \omega_{\rm o} \left[ 1- \epsilon_{\rm GW}\;G_n(\xi) 
\right]~~~~~~{\rm if~~~} a \geq a_{\rm d} \nonumber \\
\omega_{\rm b}(\xi) & = & \omega_{\rm o} \left[ 1- \epsilon_{\rm d} \;G_n(\xi) 
\right]~~~~~~{\rm if~~~} a_{\rm T} \leq a < a_{\rm d}
\end{eqnarray}
where\footnote{Note that $\epsilon_{\rm d}/ \epsilon_{\rm GW} = {\rm t}_{\rm
    GW}/ {\rm P}_{\rm o} \gg 1$, ${\rm P}_{\rm o}$ being the orbital period.} 
    $$ \epsilon_{\rm GW} =  \frac{\pi \dot{\omega}_o}{\omega_o~\omega_*}
F_n^{1/2}(\xi_{\rm out}),$$ 
$$ \epsilon_{\rm d} = \frac{\omega^2_{\rm d}}{2 \omega_* \omega_{\rm o}}
F_n^{1/2}(\xi_{\rm out}), $$
and
$$G_n(\xi) = \frac{1}{F_n^{1/2}(\xi)} = \left[\frac{\xi}{{\rm d}\theta/{\rm d}
\xi}\right]^{1/2}. $$ 
 Eq.~\ref{omega_b_r} shows that the tidal wave rotates in the secondary's
frame at a non-uniform angular speed, which decreases 
from the center outwards (Fig. \ref{fig:tide_4}).

\subsection{The total free energy}
\label{sec:totalfree}

Our results outline a new picture, which may have important consequences.
 As a result of the coupling between tidal interaction and 
orbital evolution, a radially-dependent tidal perturbation will be 
applied to the NS. 
Fluid elements at different radii within the 
NS react with different time lags at the passage of the perturbing 
primary. The element at the surface, being the slowest to react, will still be 
pointing towards an earlier position of the primary, while a central element 
will be instantaneously aligned with its current position. In other words, 
oscillations of fluid elements at different radii are out of phase 
and occur along different directions.
This condition is not consistent with hydrostatic 
equilibrium and it is not a minimum energy configuration. 

We thus anticipate that internal fluid motions will be excited, with both 
radial and azimuthal components of velocity (cfr. Goldreich \& Nicholson 1989): 
these internal motions will carry that part of energy of the 
tidal perturbation which is in excess with respect to the
equilibrium state. A complete characterization of such motions is beyond the scope of this paper. 
However, we can calculate the amount of free energy associated with them.

 To this aim, let 
us define some proper weighted average of the wave angular velocity,
$\bar{\omega}_{\rm b}$. The frame rotating at this frequency will be in
phase with the tidal wave at a particular radius. As a consequence,
an observer in that frame would see elsewhere oscillations with a
$\theta$-component of the displacement.
The amplitude of such $\theta$-component would grow with distance from the reference 
radius. The energy associated with the motion observed in this frame
represents a genuine free energy reservoir carried by the
wave. If, indeed, such motions were to be damped by some, yet unspecified, internal torque,
all rings would oscillate in phase and the wave would rotate at the 
(uniform) angular velocity $\bar{\omega}_{\rm b}$. 
The difference -- at constant total angular momentum -- between the energy in the excited state and the minimum
energy corresponding to the uniformly rotating wave is the free energy reservoir that we want to quantify.

The total angular momentum and energy carried by the wave (the tidal bulge) can be expressed as
\begin{eqnarray}
\label{totalquantities}
L_{\rm tot} & = & \int \frac{dI_b}{d\xi} \omega_b(\xi) d\xi = I_{b,0} \omega_o (I_1 - \epsilon I_2), \nonumber \\
E_{\rm tot} & = & \frac{1}{2} \int \frac{dI_b}{d\xi} \omega^2_b(\xi) d\xi = \frac{I_{b,0} \omega^2_o}{2} (I_1 - 2 \epsilon I_2 + \epsilon^2 I_3),
\end{eqnarray}
where we used eq.~\ref{eq:ib_r} to obtain
\begin{equation}
\label{dI_bdxi}
\frac{dI_b}{d\xi} = 
A^8 \frac{d I_b}{dR} =   I_{\rm b,0} \frac{d}{d\xi}
\left[\frac{\kappa_2 \xi^5}{F_n(\xi)}\right], 
\end{equation}
where $I_{\rm b,0} = 6 \pi^2 A^5 \omega^4_{\rm o}/[G^2 (1+q)^2 4 \pi \rho_{\rm c}]$.    
The integrals in eq.~\ref{totalquantities} are

\begin{eqnarray}
\label{functions}
I_1 & = & \int_0^{\pi} \frac{d}{d\xi} \left[\frac{\kappa_2 \xi^5}{F_n(\xi)}\right]~d\xi = I_b/I_{b,0}, \nonumber \\
I_2 & = & \int_0^{\pi} \frac{d}{d\xi} \left[\frac{\kappa_2 \xi^5}{F_n(\xi)}\right] G_n(\xi)~d\xi ,\nonumber \\
I_3 & = & \int \frac{d}{d\xi} \left[\frac{\kappa_2 \xi^5}{F_n(\xi)}\right] G_n^2(\xi)~d\xi.
\end{eqnarray}
The energy of a uniformely rotating wave is instead
$E_{\rm min} = (1/2) I_{\rm b} \bar{\omega}^2_{\rm b} = L^2_{\rm tot}/(2 I_{\rm b})$. 
Therefore, the free energy at a given orbital frequency is
\begin{equation}
\label{free-energy-general}
{\rm d E_{\rm f}}  \equiv  E_{\rm tot} - E_{\rm min} = \frac{I_{b,0}
  \omega^2_o}{2}\epsilon^2 \left(I_3 - \frac{I^2_2}{I_1}\right)
\end{equation}
where $\epsilon$ can be either $\epsilon_{\rm GW}$ or $\epsilon_{\rm d}$,
depending on the orbital separation under consideration. Note that this 
expression is quadratic in $\epsilon$, as expected since the minimum energy 
corresponds to $\epsilon = 0$. 

Eq. \ref{free-energy-general} can be rewritten as the product of two functions. The first one
carries the dependence on the orbital separation and goes as  
$\propto \omega^4_{\rm o} \dot{\omega}^2_{\rm o}$,
 for $a > a_{\rm d}$, and  $\propto \omega^4_{\rm o} \omega^2_{\rm d}$ for $a_{\rm T},a_{\rm c} \leq a < 
a_{\rm d}$. The second one $\propto A^{5} F_n(\xi_{\rm out})/\rho_{\rm c} \left(I_{\rm 3} - I_{\rm 2}^2/I_{\rm 1}\right)$
 depends only on the NS structure.
 A standard calculation provides the 
value of F$_n(\xi_{\rm out})$ for different polytropic eos's, while evaluating 
the integrals  in eq. \ref{functions}  is more complex. It first requires to 
derive the functional form of $\kappa_2(r)$. The details of this important 
calculation are deferred to our appendix \ref{appendixa}. Here we state the 
main result,

\begin{equation}
\label{Love-R_main}
\kappa_2(r) = \frac{1}{2} \left[\frac{2-y(r) - \Delta y(r)}{3+y+ \Delta
    y(r)}\right] \, ,
\end{equation}
where the function $y (r)$ is determined by the solution of the perturbed 
Poisson's equation for the tidal field (see appendix \ref{appendixa}). 
The quantity $\Delta y(r) = - 3 \rho(r)/\tilde{\rho}(r)$ is a measure
  of the degree of central condensation of the density profile and
  $\tilde{\rho}(r) = 3 M(r)/(4 \pi r^3)$  is the average matter density within $r$.

Tab. \ref{tab1} summarizes our main results for three different
polytropic indices corresponding to plausible NS models. Results for a 
constant-density ($n$=0) model are also given for reference.

\begin{table}
\begin{center}
\caption{Numerical values of the main structural properties of different
polytropic eos's, as explained in the text. $\tilde{{\rm F}}_n$  stands for 
${\rm F}_n(\xi_{\rm out})$.}
%{\color{red} In questa tab, e' $\tilde{F}$ NON $\sqrt{\tilde{F}}$ ? Io aggiungerei il valore di $c_1$ e ridurrei di uno o 
%due (vedi$\xi_{\rm out}$) le cifre significative}}
{\small
\begin{tabular}{c|c|c|c|c|c|c|c}
\hline
\hline
$n$ & $\xi_{\rm out}$ & $\tilde{{\rm F}}^{1/2}_n$ & $c_1$ & $\kappa_2$ & $I_1$ & $I_2$ &
  $I_3$ \\ 
\hline
0   & 2.449 & 0.58& 0.27 & 0.75 & 198.41 & 343.66 & 595.23 \\
1/2 & 2.753 & 0.43& 0.29 & 0.45 & 391.99 & 817.52 & 1712.94 \\
2/3 & 2.874 & 0.39& 0.30 & 0.37 & 493.35 & 1102.29 & 2482.19\\
1 & $\pi$   &1/$\pi$& 0.32& 0.26 & 785.01 & 2023.93 & 5299.92 \\
\hline
\hline
\end{tabular}
}
\label{tab1}
\end{center}
\end{table}
 
 We are now in the position to quantify the total free energy
that is pumped in a NS by the tides, during its late inspiral 
(eq.~\ref{free-energy-general}). 
Since the details of the orbital evolution depend on whether the primary is 
itself a NS or a BH (cfr. sec. \ref{sec:summary-scenarios}), we will discuss 
the two cases separately, beginning with NS-NS systems.
%%%%%%%%%%%%%%%%%%%%%%%%%%%%%%%%%%%%%%%%%%%%%%%%%%%%%%%%%%%%%%%%%%%%%%%
\subsubsection{Free energy in a NS-NS binary}
\label{sec:free-energy-NSNS}

Let us consider the two stages of their orbital evolution separately. 
During the GW-driven inspiral phase the total energy accumulated up to the dynamical radius $a_{\rm d}$ is
$$E^{({\rm d})}_{\rm f}  =  \int_{a_{\rm d}}^{\infty} {\rm d} E_{\rm f} = (1/2)
  \left(I_3 - \frac{I^2_2}{I_1}\right) \int {\rm d}\left(I_{\rm b,0} \omega^2_{\rm
    o} \epsilon^2_{\rm GW} \right),$$
 where we recall that $\epsilon_{\rm GW}^2 =  \left(\pi \dot{\omega}_o/\omega_o~\omega_*\right)^2
\tilde{F}_{\rm n}$, and the GW-driven orbital evolution is given by,
\be
\dot{\omega}_{\rm o} = \frac{96}{5} \frac{(GM_1)^{5/3}}{c^5} \frac{q \; \omega_{\rm o}^{11/3}}{(1+q)^{1/3}}. 
\ee
We thus obtain, 
\begin{eqnarray}
\label{free-energy-number}
E^{({\rm d})}_{\rm f}  &=& c_1^{34/3} \left(I_3 - \frac{I^2_2}{I_1}\right)
\left[ \frac{3 \pi^4 \tilde{F}^2}{\xi_{\rm out}^5}\right] \left(\frac{96\,
  c}{5}\right)^2 \times \nonumber \\
 &\times & M_* {\cal{C}}^6  \frac{(1+q)^{3}}{q^{4/3}}\;
\left(\frac{\omega_{\rm d}}{\omega_{\rm T}}\right)^{\frac{34}{3}} 
\end{eqnarray}
where $\xi_{\rm out}$ and $\tilde{F} \equiv  F_n(\xi_{\rm out})$ are given in Tab. \ref{tab1}.
Note the presence at a very high power of the ratio between the orbital frequency at
$a_{\rm d}$ and the tidal frequency. This ratio will be 
  in general much smaller than unity. Thus, a much larger amount of free energy will 
be accumulated during the subsequent stage of orbital evolution, which
  starts at $a_{\rm d}$ and ends at the contact radius $a_{\rm c}$,
\begin{eqnarray}
\label{eq:diff-en-case-A}
E^{({\rm c})}_{\rm f} & = & \int_{a_{\rm d}}^{a_{\rm c}} {\rm d} E_{\rm f} = (1/2)
  \left(I_3 - \frac{I^2_2}{I_1}\right) \int {\rm d}\left(I_{\rm b,0} \omega^2_{\rm
    o} \epsilon^2_{\rm d} \right)  \\
    &= &\frac{3 \pi^2 R_*^8 \tilde{F}^2  \xi_{\rm out}^{-5}}{4 G^2 (1+q)^2 M_*}   \left(I_3 - \frac{I^2_2}{I_1}\right)
\frac{\omega^4_{\rm d}}{\omega^2_*}~\omega^4_{\rm c} \left[1- \left(\frac{\omega_{\rm
      d}}{\omega_{\rm c}}\right)^4\right]. \nonumber
\end{eqnarray}

In order to derive numerical estimates of E$_{\rm f}$ in either case above, 
we need to specify the primary mass $M_1$, the eos
and calculate the dynamical and contact distances. 
Here, we will concentrate on three different types of systems: case A will represent two 
identical NSs with $M_* = 1.4 M_{\odot}$, $R = 12$ Km and $n=1$; case B will 
represent a mixed system where the primary is instead a massive NS, with the 
same radius but $M_* = 2 M_{\odot}$ and $n=1/2$; case C will include the same 
massive primary and a massive secondary, with $n=2/3$. 
Tab. \ref{tab2} reports the numerical values
of all quantities needed in eq. \ref{eq:diff-en-case-A} to evaluate the free 
energy pumped in each NS. 
The first is the available free
energy once $a_{\rm d}$ is reached, and is expressed in units of $10^{45}$
ergs. The second is the available free energy when the contact radius is
eventually reached, and it is in units of $10^{47}$ ergs. The corresponding 
energy in the primary can be readily obtained, by means of 
eq. \ref{eq:diff-en-case-A} and quantities in the two tables. Note however, that its  free energy 
will never exceed that in the secondary. 

\begin{table}
\begin{center}
\caption{Numerical values of the physical quantities required to calculate
the free energy in each NS (eqs. \ref{free-energy-number} and
\ref{eq:diff-en-case-A}). The three cases refer to different choices
for the binary components (see text). Different 
superscripts identify the two component stars. Numerical subscripts mean 
that the corresponding quantities are in units of the given power of 10. 
Units are in c.g.s. }
{\small
\begin{tabular}{c|c|c|c|c|c|c}
\hline
\hline
case/$q$ & $\rho^{(1)}_{\rm c,15}$ & $\rho^{(2)}_{\rm c,15}$&
$\omega_{\rm d,3}$& $\omega_{\rm c,3}$ & E$^{({\rm d})}_{\rm f,45}$ &
E$^{({\rm c})}_{\rm f, 47}$\\
\hline
A/ ~1 & 1.27 & 1.27 & 3.15 & 5.31 & $\simeq$ 4.2  & $\simeq$  11  \\
B/0.7 & 1.02 & 1.27 & 2.95 & 4.77 & $\simeq$ 4.9  & $\simeq $ 7.5 \\
C/~1  & 1.02 & 1.24 & 2.91 & 5.04 & $\simeq$  2.3  & $\simeq $ 2.7  \\
\hline
\hline
\end{tabular}
}
\label{tab2}
\end{center}
\end{table}
%%%%%%%%%%%%%%%%%%%%%%
%%%%%%%%%%%%%%%%%%%%%%%%%%%

\subsubsection{Free energy in a BH-NS binary}
\label{sec:free-energy-BHNS}
When the primary is a BH, the ISCO is the larger of the
characteristic radii, if the mass ratio is below some $q_{\rm max}$, with
$0.3 \lesssim q_{\rm max} \lesssim 0.5$, depending
on the NS eos (cfr. Fig. \ref{fig:raggiBH}). 
This corresponds to black hole masses larger than a few $M_{\sun}$.
This range includes the vast majority of BH masses in NS-BH binaries,
and we therefore focus on this case.

As for NS-NS systems, the orbital evolution is split into two phases, with the GW-driven
inspiral being overtaken by a direct plunge once $a_{\rm isco}$ is
reached. Therefore, the free energy, $E^{\rm isco}$, available down to $a_{\rm isco}$  will still
be given by eq. \ref{free-energy-number}, when we substitute 
$\omega_{\rm d}$ with $\omega_{\rm isco} \equiv \omega_{\rm o}(a_{\rm isco}) = c/a_{\rm isco} \sqrt{(1+q)/6}$. 
Values of this free energy  for four
different NSs, with a fixed M$_{\rm 1} = 7$ M$_{\rm \odot}$ 
black hole companion are reported in tab. \ref{tab3}.

For BH masses $< 14 M_{\sun}$, the merger occurs at $a_{\rm T}$ (not $a_{\rm c}$), 
which is thus the typical case for stellar mass compact binaries.
The calculation of the free energy in this plunging phase needs solving
$E^{(\rm T)}_{\rm f} =  (1/2)  \left(I_3 - (I^2_2/I_1)\right) \int_{a_{\rm isco}}^{a_{\rm T}}  {\rm d}\left(I_{\rm b,0} \omega^2_{\rm
    o} \epsilon^2_{\rm d} \right)$ where -- contrary to eq.~\ref{eq:diff-en-case-A} --  the extremes of integration are analytical functions of
the masses and $R_*$. We can thus derive an exact solution as a function of such
parameters,
\begin{eqnarray}
\label{eq:diff-en-BHNS-solved}
E^{(\rm T)}_{\rm f} = \frac{3 \pi^2 c^4_1}{2^2 6^6} \frac{\tilde{F}^2}{\xi^5_{\rm out}} \frac{M_{\rm *} c^2}{{\cal C}^5} q^4(1+q)^2 
  \hat{f}(\omega) \left(I_3 - \frac{I^2_2}{I_1}\right) & &
\nonumber \\ 
 \simeq  8 \times 10^{49} \mbox{\small{erg}}~ \frac{M_*}{1.4 M_{\sun}}
\left(\frac{0.17}{{\cal C}}\right)^{5} \left(\frac{c^2_1}{0.1}\right)^2
(1+q)^2 q^4 \hat{f}(\omega)
\end{eqnarray}
where $\hat{f}(\omega)$
$\equiv \left[1 - \left(\frac{\omega_{\rm isco}}{\omega_{\rm T}}\right)^4 \right]$.
 
 The numerical value in the second step is obtained for the usual fiducial
NS with an $n=1$ polytropic eos. Values of the free energy for four different 
NSs and a fixed M$_{\rm BH} = 7$ M$_{\rm \odot}$ are reported in
the last column of Tab. \ref{tab3}. The total energy at $a_{\rm T}$ is $\sim 10^{47}$ erg.
We note that these values depends on the BH mass via  $q^4 (1+q)^2$. 
For a $5 M_{\sun}$ BH, the available energy would be larger by a factor around $4$, while for $M_1=10 M_{\sun}$ it would be reduced by a similar factor.

\begin{table}
\begin{center}
\caption{Numerical values of the physical quantities required to calculate
the free energy pumped in the NS (eqs. \ref{free-energy-number} and
\ref{eq:diff-en-BHNS-solved}). The three cases refer to different choices
for the NS polytropic index, while M$_{\rm BH} = 7$ M$_{\odot}$ is fixed,
fixing $a_{\rm isco} \simeq 6.22 \times 10^6$ cm (M$_{\rm BH,\odot}$/7). 
Numerical subscripts mean that the corresponding 
quantities are in units of the given power of 10. Units are in c.g.s.}
{\small
\begin{tabular}{c|c|c|c|c|c|c|c}
\hline
\hline
$n$ & M$_{\rm NS,\odot}$ & $\rho_{\rm c,15}$ & $a_{\rm isco}/a_{\rm T} $& $\omega_{\rm T,3}$ & 
E$^{({\rm isco})}_{\rm f,45}$ &
E$^{({\rm T})}_{\rm f, 47}$\\
\hline
% & 2.4495 & 0.75 & 0.577 & 198.409 & 343.656 &
% 595.232\\                       
~1  & 1.4 & 1.273 & 1.42  & 3.638 & $\simeq$ 12.1  & $\simeq$ 1.6  \\
2/3 & 1.4 & 0.865 & 1.36  & 3.420 & $\simeq$  3.0  & $\simeq$ 1.3  \\
2/3 & 2   & 1.236 & 1.53  & 4.230 & $\simeq$  3.5  & $\simeq$ 1.4  \\
1/2 & 2   & 1.015 & 1.48  & 4.039 & $\simeq$  1.3  & $\simeq$ 0.8  \\
\hline
\hline
\end{tabular}
}
\label{tab3}
\end{center}
\end{table}
  
Following our calculations, (Section 7.2.1 and 7.2.2), we can conclude that in both NS-NS and BH-NS systems, 
the total free energy available is $\sim 10^{-4}$ of the total energy in the tides $E_{\rm b,T}$ 
(eq.~\ref{eq:DE_tot_tidal}). Therefore, not only the total energy involved in the 
tidal interaction is substantial but also the free energy budget. If tapped 
by additional physical processes, it may have observable consequences. We 
devote the following discussion to this possibility.
  
\section{Discussion and Conclusions}
\label{sec:discussion}
In this paper, we investigate a mechanism of tidal interaction, that appears to be
most relevant in close compact binary systems. 
We showed that, even in the limit of zero viscosity, a mutual tidal torque is
maintained successively by the GW-driven orbital evolution and by the final dynamical instability before merger. 
The orbital energy transferred to the bulge is 
quite substantial: $10^{-3} - 10^{-2}$ of the secondary binding energy.

This result is in agreement with 
Goldreich \& Nicholson (1989). They consider an inviscid and
self-gravitating fluid, subject to an external potential, which rotates rigidly
at a given frequency.
They find that, in their case, there is no secular variation
of the angular momentum of fluid particles. 
Indeed, we showed that, for a constant orbital separation, an immediate synchronization
occurs after which the torque vanishes. The torque is only sustained by the orbital shrinking,
and it is thus related to secular evolution of the total orbital angular momentum

We then studied the radial structure of the rotating quadrupolar perturbation, excited
within the secondary star by this {\em orbital evolution-driven} torque. We concluded that, quite generally,
it inherits the differential structure of the tidal torque. In the secondary frame, the perturbation
can be described as a wave excited by the motion the the primary. The crest of the wave at different radii
rotates at different rates: while the inner regions promptly follow the motion of the primary, the outer regions
lag progressively behind. 

This is not a minimum energy configuration, that would correspond to a
 uniformly rotating quadrupole.
For different binary systems, we calculated the free energy associated with this excited state, and found that
typically $ \sim 10^{-3}-10^{-4}$ of the total tidal energy should be removed
from the bulge in order to reach the equilibrium state.
This free energy reservoir corresponds to $10^{47}-10^{48}$ erg.
The most favourable case ($10^{48}$ erg) corresponds to an equal mass, double NS system with $n=1$.

A few important considerations lead us to conclude that our estimates
represent a {\em lower limit} on the total free energy. First, we adopted 
the most conservative description of the plunging phase, which
assumes that the lag angle remains constant during the free 
fall (sec. \ref{sec:diff-tidal-wave}).
In fact, this angle is more likely to increase as the orbital separation
  decreases, thus providing a factor of a few more free energy than we estimated. For instance for our BH-NS
systems (see Tab. \ref{tab3}), the gain would be a factor $(a_{\rm isco}/a_{\rm T})^3  \approx 2.5-3.6$,
  if the lag angle were to grow linearly with
  $\omega_{\rm o}(a)$, as suggested by its definition
  (eq. \ref{alpha_dyn_R}). 
Second, our treatment is newtonian and, third, linear in the lag angle. Both these
features are known to underestimate the effects of forces and
torques, thus the free energy generation, close to coalescence, where most of the energy is produced.

We expect this free energy to be transferred to internal motions, via excitation of low-frequency g-modes well before
coalescence and of more effective f-modes, just prior to coalescence, when the
orbital frequency approaches the fundamental one, $\omega_*$. We defer to a
future work a detailed study of this problem, but it is reasonable
to expect that the azimuthal motions excited by the
differentially rotating quadrupolar perturbation  will largely inherit its differential structure.

This suggests that a pre-existing magnetic field in the
NS core might get significantly sheared and amplified. The free energy in the excited
  internal motions grows extremely fast with orbital frequency and, at the tidal radius, it would 
correspond to a - mostly azimuthal - magnetic field $B_{\phi} \approx
10^{15}$ G, permeating the whole NS volume. As mentioned above, we are probably underestimating the free energy 
content and thus the magnetic field, which nevertheless is already as strong as in the most magnetized objects known. 
In fact, this dynamically-driven tidal torque may be able to ``recycle" an old NS into a magnetar.

The question is: would the free energy be tapped fast enough to build such a strong 
field component by the time the tidal radius is reached?
A detailed analysis of this scenario is clearly beyond our scope here, as it
would require solving for the fluid motions and for their coupling to the
magnetic field. However, there is an argument that suggests that this may be possible. 
In a pure $\Omega$-dynamo the azimuthal field would grow linearly and 
become as strong as the initial poloidal one, {\em before} the dynamical 
radius is reached ($a > a_{\rm d}$). Thus, where most of the free energy 
is generated ($a \le a_{\rm d}$), we might expect the linear
  approximation to break down and field growth to proceed in a non linear 
regime. A growth timescale shorter by a factor of $5-10$ than in the
linear regime
would be already sufficient to tap the entire free energy reservoir. As 
a matter of fact, when the azimuthal field carries most of
the magnetic energy, conditions are such that shear instabilities may be 
excited by the differential rotation, (e.g. Balbinski 1985; Luyten 1990; 
Shibata, Karino \& Eriguchi 2002; Watts, Andersson \& Jones 2005), and the 
resulting turbulent fluid motions would lead to a much faster growth of this 
field component.

The above discussion indicates possible measurable consequences during 
the final coalescence of a binary system containing a NS. 
If this non-linear phase happens, such a strong magnetic field may 
manifest itself in a magnetar-like flare at the moment the NS is tidally
disrupted, as it may actually have been observed already (cfr. Troja et al. 2010).
Alternatively, the magnetic field may permeate in a toroidal fashion the NS
debris, that will form an accretion disc around the remnant of the
merger. This can favour the magnetic production of jets, like it is thought to happen for short
$\gamma$-ray bursts.

\appendix

\section{The radial profile of the tidal Love number}
\label{appendixa}
The tidal Love number $\kappa_2$ for NSs has been calculated 
in fully relativistic models (Hinderer 2008, D'amour \& Nagar 2009,   Poisson \& Binnington
2009, Postnikov et al. 2010), allowing for different choices of the 
equation of state. Building on 
these works, we sketch here a simple derivation of its radial
profile within the NS, $\kappa_2(r)$, in the newtonian limit. 
We start from the relativistic formulation (Hinderer 2008 and references therein), 
where the spacetime metric is written as the sum of a (known) background and a 
perturbation, in axial symmetry.  The metric perturbation contains the
  Laplacian of the tidal field of the primary, $\nabla^2 \psi_{\rm T}$ (cfr. sec. \ref{sec:Nt_integral}), and the
induced quadrupole moment of the secondary, $Q_{22}$, as leading orders terms. 
Their ratio defines the Love number $\kappa_2$ of the
star (Hinderer 2008 and references therein)
\begin{equation}
\label{eq:def-lovenumber}
\kappa_2 \equiv \frac{3}{2}\frac{G}{R_*^5} \frac{Q_{22}}{\nabla^2 \psi_{\rm T}}.
\end{equation}
An appropriate choice of the reference frame and gauge conditions, 
reduces the
Einstein's equations for the metric perturbation to a single,
second-order differential equation for  the radial function, $H(r)$, which describes
the $tt$-component of the metric perturbation.
%

%%%%%%%%%%%%%%%%%
\begin{figure}
\includegraphics[width=0.48\textwidth]{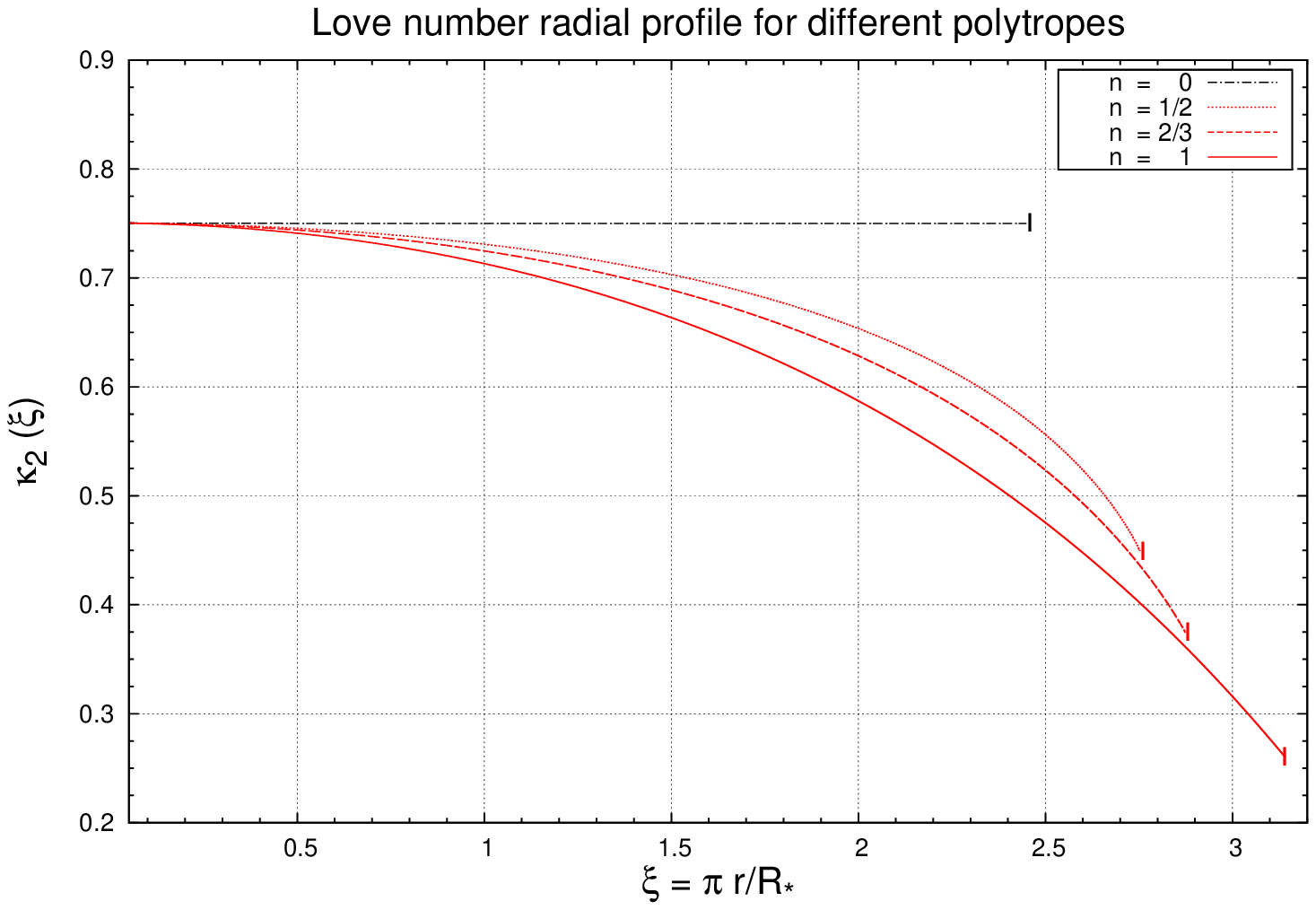}
\caption{The radial profile of the NS Love number, $\kappa_2(\xi)$, for 
four different polytropic eos's,  
%As figure \ref{fig:tide_3} right panel, but for the Love number $\kappa_2$ 
calculated through eq. \ref{Love-R}.}
\label{fig:tide_5}
\end{figure}
%%%%%%%%%%%%%%%%%%%

Solutions for $H(r)$ are then found separately inside the NS, (given
a specific equation of state), and in the vacuum outside it. 
The outside metric is matched asympotically to the Schwarzschild solution,
while a regular behaviour as $r \rightarrow 0$ is imposed to the interior solution. 
The two solutions must eventually be matched at the NS surface and 
this requires that both $H(r)$ and $H'(r)$ be continuous across it 
(cfr. Hinderer 2008, Postnikov et al. 2010). This matching determines the coefficient $\kappa_2$, 
as a function of the value of the interior solution at $r=R_*$.

By taking the newtonian limit of the relativistic solution, i.e. the limit for small stellar
compactness, and introducing the variable $y(r) = r H'(r)/H(r)$, this
procedure eventually recovers the classical result (cfr. Brooker \& Olle 1955)

\begin{equation}
\label{Love}
\kappa^{N}_2 = \frac{1}{2}\left(\frac{2-y}{3+y}\right) ,
\end{equation}
where $y$ here takes the value given by the interior solution for $H(r)$ at $r=R_*$.
To find the run of $\kappa_2$ with radius, 
we need to cut a smaller sphere of radius $r < R_*$ within the tidally
  deformed NS, and remove the outer ring \textit{without changing the matter
    distribution within.} Then, we can calculate the Love number of the smaller sphere, $\kappa_2(r)$,
by matching the interior solution with the exterior solution at radius $r$. 
However, since the matter distribution within the smaller sphere is unchanged,
it is still determined by the full polytropic model. Hence,  
$\rho(r) \neq 0$ and the matter density has a discrete jump at the surface of
the sphere,
from $\rho (r)$ to zero. In presence of a density
discontinuity, the variable $y(r)$ itself will have a discrete jump, $\Delta
y$, which is determined by the new matching condition at radius $r$.

Following Postnikov (2010), it is possible to write this jump as
\begin{equation}
\label{eq:define-deltay}
\Delta y = 3~\frac{\rho(r^+) - \rho(r^-)}{\tilde{\rho}(r)},
\end{equation}
where $r^+$ and $r^-$ indicate respectively an \textit{external
    point} and an \textit{internal point}, infinitesimally close
  to the surface of the sphere. In our case, $\rho(r^+) = 0$, $\rho(r^-) =
\rho(r)$. Finally,
$\tilde{\rho}(r) = 3 M(<r)/(4 \pi r^3)$ is the average matter density
within radius $r$ . Therefore, the appropriate function to insert into
(eq. \ref{Love}), is
$y(r)+\Delta y$, with $\Delta y$ given by eq. \ref{eq:define-deltay}.
%- 3\rho(R)/\tilde{\rho}(R)$,{\bf  where $y(R)$ 
%represents the value of $y$ at a radius $R$ for the polytropic model. questa frase non e' chiara}
With this procedure we obtain,
\begin{equation}
\label{Love-R}
\kappa_2(r) = \frac{1}{2} \left[\frac{2-y(r) - \Delta y(r)}{3+y+ \Delta y(r)}\right],
\end{equation}
which gives expression (\ref{Love}) for $r \rightarrow R_*$ and goes to 3/4 (the value of a constant density sphere) as $r \rightarrow 0$.
These are the correct limits\footnote{The expression for a sphere of uniform density is given by Brooker \& Olle (1955) in terms of the polytrope 
of index $n=0$ and equals $3/4$. Postnikov (2009) writes that $\kappa_2 = 0$ for this specific polytrope, because that author does not account for the
correction due to the density jump that will inevitably appear at the boundary of this model. Adding this correction, as we have described it here, 
gives the value of $3/4$ also for the calculation by Postnikov (2010).} for the function $\kappa_2(r)$.

A generalization of this procedure to the relativistic case is conceptually straightforward.

\section*{Acknowledgements}
We acknowledge very useful discussions with Y. Levin, P. Armitage, R. Sari and T. Piran.
SD acknowledges support from ERC advanced research grant ``GRBs".

\label{lastpage}

\begin{thebibliography}{}

\bibitem[Alexander(1973)]{1973Ap&SS..23..459A} Alexander, M.~E.\ 1973, \apss, 23, 459 

\bibitem[Andersson(2007)]{2007Ap&SS.308..395A} Andersson, N.\ 2007, \apss, 308, 395 

\bibitem[Balbinski(1985)]{}Balbinski E., 1985, A\&A, 149, 487

\bibitem[Bildsten 
\& Cutler(1992)]{1992ApJ...400..175B} Bildsten, L., \& Cutler, C.\ 1992, \apj, 400, 175


\bibitem[Bini, Damour \& Faye(2012)]{} Bini, D., Damour, T., \& Faye, G. 2012, Ph.Rv.D., accepted (arXiv:1202.3565)

\bibitem[Brooker \& Olle(1955)]{1955MNRAS.115..101B} Brooker, R.~A., \& Olle, T.~W.\ 1955, \mnras, 115, 101


\bibitem[Chandrasekhar(1933)]{1933MNRAS..93..449C} Chandrasekhar, S.\ 1933, 
\mnras, 93, 449 

\bibitem[Clark \& Eardley(1977)], Clark, J.P.A., \& Eardley, D.M. \ 1977, \apj, 215, 311 

\bibitem[Damour 
\& Nagar(2009)]{2009PhRvD..80h4035D} Damour, T., \& Nagar, A.\ 2009, \prd, 80, 084035 

\bibitem[Darwin(1879)]{1879Obs.....3...79D} Darwin, G.~H.\ 1879, The 
Observatory, 3, 79

\bibitem[Goldreich 
\& Nicholson(1989)]{1989ApJ...342.1075G} Goldreich, P., \& Nicholson, P.~D.\ 1989, \apj, 342, 1075 

\bibitem[Hinderer(2008)]{2008ApJ...677.1216H} Hinderer, T.\ 2008, \apj, 
677, 1216

\bibitem[Hut(1981)]{1981A&A....99..126H} Hut, P.\ 1981, \aap, 99, 126 

\bibitem[Kochanek(1992)]{1992ApJ...398..234K} Kochanek, C.~S.\ 1992, \apj, 
398, 234 

\bibitem[Kopal(1959)]{1959cbs..book.....K} Kopal, Z.\ 1959, The 
International Astrophysics Series, London: Chapman \& Hall, 1959 

\bibitem[Kopal(1968)]{1968Ap&SS...1..179K} Kopal, Z.\ 1968, \apss, 1, 179

\bibitem[Lai, Rasio \& Shapiro(1994a)]{} Lai, D., Rasio, F.A., \& Shapiro S.~L. \ 1994, \apj, 420, 811

\bibitem[Lai, Rasio \& Shapiro(1994b)]{} Lai, D., Rasio, F.A., \& Shapiro S.~L. \ 1994, \apj, 437, 742

\bibitem[Lai \& Shapiro(1995)]{} Lai, D., \& Shapiro S.~L. \ 1995, \apj, 443, 705

\bibitem[Lattimer 
\& Prakash(2001)]{2001ApJ...550..426L} Lattimer, J.~M., \& Prakash, M.\ 2001, \apj, 550, 426

\bibitem[Lattimer 
\& Prakash(2006)]{2006NuPhA.777..479L} Lattimer, J.~M., \& Prakash, M.\ 2006, Nuclear Physics A, 777, 479 

\bibitem[Lebovitz 
\& Zweibel(2004)]{2004ApJ...609..301L} Lebovitz, N.~R., \& Zweibel, E.\ 2004, \apj, 609, 301 

\bibitem[Love(1909)]{1909MNRAS..69..476L} Love, A.~E.~H.\ 1909, \mnras, 69, 
476 

\bibitem[Luyten(1990]{} Luyten P.J., 1990, \mnras, 242, 447

\bibitem[Mizerski 
\& Bajer(2009)]{2009JFM...632..401M} Mizerski, K.~A., \& Bajer, K.\ 2009, Journal of Fluid Mechanics, 632, 401 

\bibitem[Parker(1955)]{1955ApJ...122..293P} Parker, E.~N.\ 1955, \apj, 122, 
293 

\bibitem[Pannarale et al.(2011)]{2011ApJ...727...95P} Pannarale, F., 
Tonita, A., \& Rezzolla, L.\ 2011, \apj, 727, 95 

\bibitem[Poisson 
\& Binnington(2009)]{2009APS..APRJ11001P} Poisson, E., \& Binnington, T.\ 2009, APS Meeting Abstracts, 11001 

\bibitem[Postnikov et al.(2010)]{2010PhRvD..82b4016P} Postnikov, S., 
Prakash, M., \& Lattimer, J.~M.\ 2010, \prd, 82, 024016 

\bibitem[Shibata et al.(2002)]{} Shibata M., Karino S., \& Eriguchi Y., 2002, \mnras, 334, L27

\bibitem[Taniguchi et al.(2008)]{2008PhRvD..77d4003T} Taniguchi, K., 
Baumgarte, T.~W., Faber, J.~A., \& Shapiro, S.~L.\ 2008, \prd, 77, 044003 

\bibitem[Thorne(1998)]{1998PhRvD..58l4031T} Thorne, K.~S.\ 1998, \prd, 58, 
124031 

\bibitem[Troja et al.(2010)]{2010ApJ...723.1711T} Troja, E., Rosswog, S., 
\& Gehrels, N.\ 2010, \apj, 723, 1711 


\bibitem[Watts et al.(2005)]{}Watts, A.~L., Andersson, \& N., Jones, D.~I. 2005, \apj, 618, 37

\bibitem[Zahn(1977)]{1977A&A....57..383Z} Zahn, J.-P.\ 1977, \aap, 57, 383 

\end{thebibliography}
\end{document}